\documentclass[aps,nofootinbib]{revtex4-1}

\usepackage[final]{graphicx}
\usepackage[font={small}, labelfont=bf, justification=centering]{caption}[2007/01/07]
\usepackage{natbib}
\usepackage{amsmath}
\usepackage{amsbsy}
\usepackage{caption}
\usepackage{breqn}
\usepackage{amssymb}
\usepackage{bm}
\usepackage{mathtools}
\usepackage{units}
\usepackage{cleveref}
\usepackage{color,soul}
\usepackage{xr}
\usepackage[section]{placeins}
\usepackage{algorithm}
\usepackage{algpseudocode}

\usepackage{changes}
\definechangesauthor[name={Per cusse}, color=orange]{per}
\setremarkmarkup{(#2)}

\makeatletter
\def\BState{\State\hskip-\ALG@thistlm}
\makeatother

\DeclareMathAlphabet\mathbfcal{OMS}{cmsy}{b}{n}

\usepackage{mymacros}

\usepackage{xargs}                      
\usepackage{fullpage,lmodern}
\usepackage[final]{showlabels,rotating} 

\graphicspath{ {figs/} }

\begin{document}

\title{Systematic derivation of hybrid coarse-grained models}

\author{Nicodemo \surname{Di Pasquale} \footnote{Alphabetical order}}
\affiliation{Department of Mathematics$,$ University of Leicester$,$ University Rd$,$ Leicester LE1 7RH$,$ UK}
\email[Email: ]{ndp8@leicester.ac.uk}
\author{Thomas \surname{Hudson}}
\affiliation{Warwick Mathematics Institute$,$ University of Warwick$,$ Coventry CV4 7AL$,$ UK} 
\email[Email: ]{t.hudson.1@warwick.ac.uk}
\author{Matteo \surname{Icardi}}
\affiliation{School of Mathematical Sciences$,$ University of Nottingham$,$ Nottingham NG7 2RD$,$ UK}
\email[Email: ]{matteo.icardi@nottingham.ac.uk}
\date{\today}

\begin{abstract}
  Molecular dynamics represents a key enabling technology for applications ranging from biology to the development of new materials. However, many real-world applications remain inaccessible to fully-resolved simulations due their unsustainable computational costs and must therefore rely on semi-empirical coarse-grained models. Significant efforts have been devoted in the last decade towards improving the predictivity of these coarse-grained models and providing a rigorous justification of their use, through a combination of theoretical studies and data-driven approaches. One of the most promising research effort is the (re)discovery of the Mori-Zwanzig projection as a generic, yet systematic, theoretical tool for deriving coarse-grained models. Despite its clean mathematical formulation and generality, there are still many open questions about its applicability and assumptions.
	In this work, we propose a detailed derivation of a hybrid multi-scale system, generalising and further investigating the approach developed in [Espa\~{n}ol, P., EPL, 88, 40008 (2009)]. Issues such as the general coexistence of atoms (fully-resolved degrees of freedom) and beads (larger coarse-grained units), the role of the fine-to-coarse mapping chosen, and the approximation of effective potentials are discussed.
	The theoretical discussion is supported by numerical simulations of a monodimensional nonlinear periodic benchmark system with an open-source parallel Julia code, easily extensible to arbitrary potential models and fine-to-coarse mapping functions.
	The results presented highlight the importance of introducing, in the macroscopic model, non-constant fluctuating and dissipative terms, given by the Mori-Zwanzig approach, to correctly reproduce the reference fine-grained results, without requiring \emph{ad-hoc} calibration of interaction potentials and thermostats.
\end{abstract}

\maketitle


\section{Introduction}

Molecular Dynamics simulations (MD) have become a standard tool in many applied research areas, such as the study of biological molecules \citep{Karplus2002}, soft matter \citep{Barrat2010} and, more generally, in condensed matter physics \citep{Li2017a, Daw1993}. Despite their widespread use, fully atomistic MD simulations are heavily constrained in terms of system size and simulation time, due to the computational power required to generate trajectories which are long enough to exhibit characteristic system behaviour. In fact, while solving the equations of motion for few atoms represents a simple task, calculating solutions for interesting systems such as proteins, involving hundreds of thousands of atoms and billions of time steps, rapidly becomes intractable. The key numerical bottleneck in large-scale simulations is the evaluation of forces, which are generally both numerically stiff and expensive to compute. Moreover, the precise evolution of the full system often contains irrelevant information. In the case of a protein molecule in water, a large part of the overall computation time is used to compute the evolution of water molecules, while the conformation of the protein is usually the most interesting aspect of the simulation, and may be described by relatively few variables.

Another problem often encountered in the simulation of long chain-like molecules is that as the size of the system increases, the relaxation time required to approach equilibrium becomes comparable with the maximum trajectory length achievable in simulations. In order to overcome these limitations, Coarse-Grained (CG) models have been developed alongside atomistic ones. Several `coarse-graining' techniques have been established (both systematically and empirically), allowing faster and larger Molecular Dynamics (MD) simulations than those possible with fully atomistic simulations. These typically reduce the number of degrees of freedom by grouping a number of atoms into a single particle, referred to as a `bead'. The size of such beads can range from a few atoms to entire molecules.

In this work, we focus on the Mori-Zwanzig (MZ) formalism, which is a mathematical framework which allows for systematic treatment of coarse-graining approaches. Coarse-graining an MD system requires choosing a mapping operator which describes the reduced degrees of freedom. This choice leads to an effective potential, which governs the motion of the coarse-grained variables. While this effective potential preserves various equilibrium properties of the system, it fails to preserve dynamical properties, which require a proper treatment of the fluctuations in the system. The MZ formalism provides a framework which allows us to do exactly that, and can therefore inform a choice of effective dynamics which more closely replicate the dynamical properties of the coarse-grained system.

The main objective and novelty of this work is to propose a general framework for hybrid models, explore the implications of many of the various approximations required to use the MZ formalism in practice, and provide a practical open-source implementation as a testing platform to understand and quantify the accuracy of various modelling assumptions. The methodologies and the results, obtained here for a simple one-dimensional test-case, can form the guidelines to apply the coarse-graining to more realistic chemical systems. With this in mind, below and in the conclusions we provide a summary of the specific issues and lessons learnt, while extensive computational studies on multidimensional systems are left for future works. 

The the paper is organised as follows: after a literature review of the existing methods (\cref{Sec:review}), and the definition of the notation used throughout the paper (\cref{Sec:Def}), we present the general formal equations which arise upon coarse-graining using the MZ formalism in \cref{sec:zwanzig}. We provide a detailed derivation to make this approach more accessible to non-specialists, including a discussion of all relevant derivation steps in the main text (and additional details in the Supplemental Material, SM). In \cref{sec:analysis}, we then present an analysis of the equations in their final form, while in \cref{sec:discussions} some of the approximations usually considered in literature are discussed. Our numerical results are presented in \cref{Sec:Res}, where these equations are implemented for a simple test case, which allows us to discuss practical algorithmic details. In \cref{Sec:Conclusion} we draw before drawing some conclusions and present the outlook for future studies.
Before proceeding with this programme, we outline below the most important features of the MZ formalism, with a particular focus on the consequences for practical coarse-grained MD simulation.


\subsection{Summary of results and practical consequences}

The MZ formalism treats the chosen coarse-grained variables as observables of the underlying `true' dynamics, which leads to the derivation of the MZ equations \tcref{eq:mz}. These govern the coarse-grained dynamics for the variables of choice, without approximations, resulting in a system that is no simpler to solve than the full dynamics. The value of this formalism, however, lies in separating out different contributions to the evolution of the coarse-grained variables, each of which can be assigned heuristic meaning, and can inform an approximation strategy. In particular, the terms in this equation fall into three important groups, which we discuss in turn.

The first term in \tcref{eq:mz}, after appropriate manipulations, corresponds in \tcref{Eq:DevB} to a derivative of the \emph{effective potential} (defined rigorously in \tcref{Eq:EntrInt}), representing the \emph{mean force} between beads and between atoms and beads. In \tcref{Sec:effPot}, we argue that this quantity can be approximated by many methods, including various well-known techniques available in the literature such as Iterative Boltzmann Inversion, Force Matching or the Relative Entropy method. In all cases, the force experienced by a bead in a hybrid simulation is related to the effective potential via \tcref{Eq:VeffB}, and each of these methods seek to approximate the resulting coarse-grained forces, i.e. derivatives of $\Veff$. In the numerical experiments carried out here, we use a direct way to approximate the effective potential that allows us to concurrently approximate the other terms in the MZ equation. In its current form, the direct method chosen is easily applied to a system of this size, but unlike other methods available in the literature, might not be appropriate from a computational point of view for more complex systems.
Approximating the effective potential necessarily relies on some assumptions to simplify the calculation of this highly complex multidimensional integral \tcref{Eq:EntrInt}. These are stated explicitly and discussed in \tcref{Sec:ApproxVeff}.

Regardless of the method used to approximate the effective potential, the MZ formalism provides a framework that paves the way towards a more quantitative estimation of the errors committed.

The remaining terms in \tcref{eq:mz} (and its more explicit form \tcref{Eq:DevB}), usually neglected in applied coarse-graining approaches, correspond to \emph{memory effects} (represented by time convolution of the friction matrix \tcref{Eq:fricmatrix}), and a \emph{fluctuating force}.
These terms naturally appear as a way to implicitly reintroduce the effects of the degrees of freedom that have been averaged out, similar to an external `bath'.
Their overall effect is to slow beads down to retrieve the correct dynamical properties of system. In particular, the fluctuating force term represents those forces which cause deviations from the `mean force'. 
Sampling fluctuating forces compatible with given coarse-grained degrees of freedom requires the resolution of a constrained (or orthogonal) dynamics. These also appear in the memory term which is computed as their time covariance, \tcref{Eq:fricmatrix}, and represents the effects of the past history of the system on its evolution.
When this covariance decays sufficiently fast in time, the memory term can be conveniently approximated as a simple friction term (see \tcref{Eq:convFlucF}), and the fluctuating force can be approximated as a white noise.
This desirable property makes the system Markovian, \tcref{Sec:MarkovApprox}, and this approximation is generally appropriate when there is a separation of timescales between the coarse-grained variables and those neglected.

In summary, from the theoretical derivation and the numerical results, we note the following practical consequences of our results:
\begin{itemize}
\item So far, most coarse-grained approaches, have focused on the approximation and computation of the effective potential. While its central role is recognised here, we argue that a proper understanding of the additional fluctuations and memory terms can avoid ad-hoc fitting and modifications of the potential, keeping it uniquely defined for a given mapping.
\item We observe that, to retrieve satisfactory equilibrium and dynamical properties, a simple Langevin dynamics, with constant friction and diffusion coefficients, is not appropriate even in the particularly simple numerical example we consider.
\item  A better description of static properties are obtained when the friction and diffusion coefficients are parametrised (similarly to the effective potential) with respect to the system configuration (e.g., positions).
\item The preservation of more complex dynamical properties instead requires the strict validity of the time-scale separation assumption or else a proper parametrisation and implementation of the memory term.
\end{itemize}

\section{State of the art}
\label{Sec:review}
Any coarse-graining procedure must prescribe a means by which to compute interactions between beads, and numerous variants exist in the literature \citep{Carbone2013,Brini2013}, roughly divided between structural and thermodynamic models. The former class includes models where interactions among beads are obtained by molecular structures coming from atomistic simulations. In the Iterative Boltzmann Inversion \citep{Muller-Plathe2002}, an iterative procedure progressively optimises a tabulated potential in order to reproduce a known observed quantity, such as the Radial Distribution Function (RDF); a similar approach, that aims at obtaining interactions among beads using the RDF, through Monte Carlo simulations is the Inverse Monte Carlo \citep{Lyubartsev1995}. The force-matching model \citep{Izvekov2004,Izvekov2005,Noid2008} uses a least squares minimisation over forces sampled from finer scale models. In the Relative Entropy framework \tcitep{ScottShell2008} the CG interactions are obtained by minimising an entropy function which represents the overlap between two molecular ensembles, namely the CG and the underlying atomistic one. The second class of thermodynamic models are those where thermodynamic data, either from atomistic simulations or experiments, are used to obtain interactions among beads. The approach is to decide a fixed functional form and then choosing parameters to reproduce empirical observables. The parameters thus obtained can be optimised to be valid for a generic class of chemical species as in MARTINI force field \citep{Marrink2007} or be specific for each type of molecules \citep{Shinoda2007}. In all the cases these calculated interactions represent an appropriate average of the interactions felt by the group of atoms in the full atomistic model that constitute the bead.

Most CG approaches, however, suffer from a major flaw: even if static equilibrium properties are preserved in the passage from atomistic to CG systems, the same is not true of the dynamical properties. This effect arises from the fact that neglecting degrees of freedom during the coarse-graining results in the elimination of high-frequency fluctuating components of the force \citep{Li2014}. Without these, the CG system exhibits an artificially `accelerated' dynamics which can, however, be corrected to achieve more accurate recovery of dynamical observables by adding appropriate friction terms. An example of a method to derive these terms is given in Izvekov and Voth \citep{Izvekov2006a}, where friction coefficients are derived in the framework of the multi-scale coarse-graining method \citep{Izvekov2005}.
In the framework of nonlinear equations, neglecting the friction terms mentioned above has been referred to as the \emph{first optimal description} \citep{Chorin2002,Chorin1998}, and in effect, this choice evolves the mean value of the CG variables only.  It has been shown that the first optimal description is accurate only for short times \citep{Hald2001} and is therefore clear that for sampling applications, this is insufficient to appropriately predict many observables of interest.

A second necessary limitation of coarse-grained models is the inevitable price paid for the speed-up of simulations: a loss of information about the model caused by the procedure of grouping atoms into beads. If coarse-graining is carried out incoherently, the limited level of detail offered by CG models may not be enough to correctly predict the system evolution, for example due to phenomena observed only at atomistic level ruled out by the coarse-graining procedure. An example of one such phenomenon is hydrogen bonds, which require special treatment to be considered  \citep{Karimi-Varzaneh2008} in a CG system. Another example arises in the simulation of coarse-grained polymer melts, where the entanglement regime may not be properly described because of the ``chain-crossability'' resulting from the soft CG potentials \citep{Masubuchi2014}.
These limitations have led to the development of hybrid atomistic/CG models which in principle should combine both a lower computational cost than fully atomistic simulations, along with improved accuracy over other CG schemes. Examples of such hybrid approaches include the recent works of \citet{Rzepiela2011}, where massless virtual sites which mediate interaction between atoms and beads were used for butane and dilanine in water; and \citet{DiPasquale2012a, DiPasquale2014} for polystyrene  and polyethylene melt, in which the atomistic and CG descriptions were seamlessly embedded in the same molecule. This represents an important step towards a computable and yet accurate multi-scale framework. However, it was shown that when the dynamic in hybrid models of complex systems such as polymers is considered, other effects arise. In melt, poly-ethylene chains seem to move according either to Rouse or entanglement regime depending on their resolution (i.e. on the ratio atom/beads) \citep{DiPasquale2017}. Therefore, a rigorous treatment of these models becomes essential.

As mentioned above, the CG approach studied here is the Mori-Zwanzig projection formalism. This is a promising approach to better understand and derive generic coarse-grained models, and due to its generality, can be applied to a very wide class of dynamical models than many standard CG techniques, including multi-scale (or hybrid) CG/atomistic models. In 2009, Espa{\~n}ol \citep{Espanol2009} proposed a derivation of a hybrid atomistic/CG model based on the Mori-Zwanzig operator formalism, and the equations of motion were derived for a generic polymer chain represented by its centre of mass and a single, fully atomistic, molecule. This work was based on previous studies \citep{Hijon2009,Kinjo2007} which proposed a derivation of a CG model in terms of equations of motion for beads starting from atomistic systems by using the Mori-Zwanzig (MZ) projection \citep{Zwanzig1961,Mori1965} operator. Guenza \citep{Guenza1999} proposed the use of projection operator technique for simulations in dense melts, and this approach was later derived from first principles \citep{Lyubimov2011}. At least formally, the MZ projection operator allows the reduction of the dimensionality of the system under analysis in a mathematically optimal way. One of the first applications of the MZ projection operator to CG dynamics is reported in \citet{Givon2004} where the problem of the scale separation is also addressed. 

The Mori-Zwanzig formalism, as it will become clear later, has two intrinsic advantages: First, given an arbitrary and generic mapping between atomistic and CG variables, it allows a rigorous derivation of equations for the exact evolution of CG variables, including terms which may be identified as fluctuating forces and a dissipative memory (consistent with the fluctuation-dissipation theorem). Secondly, under the hypotheses used to derive the equations, it provides an explicit, rigorous means by which to compute the CG interactions, and a framework in which to quantify and progressively improve the accuracy if desired, which is the ultimate goal of any coarse-graining methodology.
Despite being given in an explicit form, in practice the different terms in the CG equations of
motion derived via the MZ formalism are not easy to compute in a realistic scenario, and to date,
it appears no systematic way of selecting an optimal CG strategy has been developed. In recent
years, there has been a significant effort towards addressing these issues and thus to render
simulations using this theoretical tool more practicable. It is worth mentioning here that the MZ formalism leads to equations with the same structure \citep{Hijon2009} as those of the
DPD equations \citep{Groot1997}. With regards to the particularly cumbersome estimation of the memory terms, \citet{Li2014, Li2015, Li2017} proposed a way to introduce them into a CG simulation by using non-Markovian DPD.

Notwithstanding these difficulties, the MZ formalism represents an appropriate formal framework for systematic coarse-graining and for better understanding the various hypotheses behind the derivation of other CG models. In this work we present a full bottom-up derivation of a general CG model appropriate for application to MD simulations, and generalise the work of \citet{Espanol2009} by considering the presence of an arbitrary number of beads and atoms in the same molecule, rather than assuming that each molecule is represented as a single bead. Atomistic/CG models \citet{Espanol2009,DiPasquale2012a,Rzepiela2011} are usually referred to as ``hybrid''. More generally, we can view them as a broad class of models in which the fully coarse-grained model and the fully atomistic models are limiting cases, with all the possible choices and combinations of atoms/beads in between. This clearly has many advantages but induces the very challenging task of choosing the appropriate (optimal) description. We believe that, a better theoretical and practical understanding of the MZ formalism represents the starting point to quantify the approximations used in CG simulations. A few other details of the derivation and modelling assumptions, partly overlooked in previous literature, are also discussed in the following. These, together with the unavoidable further approximations introduced in the practical implementations, are discussed for a simple test case (Lennard-Jones periodic chain), solved and coarse-grained through an efficient parallel open-source implementation in \texttt{Julia v0.6} \citep{Bezanson2017,thomas_hudson_2018_1226590}.

\section{Definitions and notation}\label{Sec:Def}
Our starting point is a full dynamical system composed by $\nfg$ atoms, to which we refer as the \emph{fine-grained} (FG) system. The term \textit{particle} is used to indicate either an atom or larger group of atoms, called beads.
We will assume that the chemical structure of the system is fixed (i.e. no reactions occur) and the connectivity of the atoms is specified at the outset, allowing us to fix a global index for the particles in the FG system. Throughout this paper, FG particles are referred to with lower-case indices, e.g. $i,j$.
Correspondingly, phase space variables characterising the degrees of freedom of particles in the FG system will be denoted with lower case bold
letters, e.g. $\bfr$ represents their positions, with $\bfp$ representing their momenta
In the same spirit, we suppose that the coarse-grained (CG) description of the same system is composed of $\ncg$ particles. Each FG particle is mapped to a fixed (unique) particle in the CG system. CG particles will be indexed with capital letters, e.g. $I,J$. Phase space variables characterising the degrees of freedom of particles in the CG system will be denoted with capital
letters, e.g. $\bfR$ represents the position of the CG particles while $\bfP$ represents their momenta.

\begin{figure}[t]
    \centering
    \includegraphics[width=0.4\textwidth]{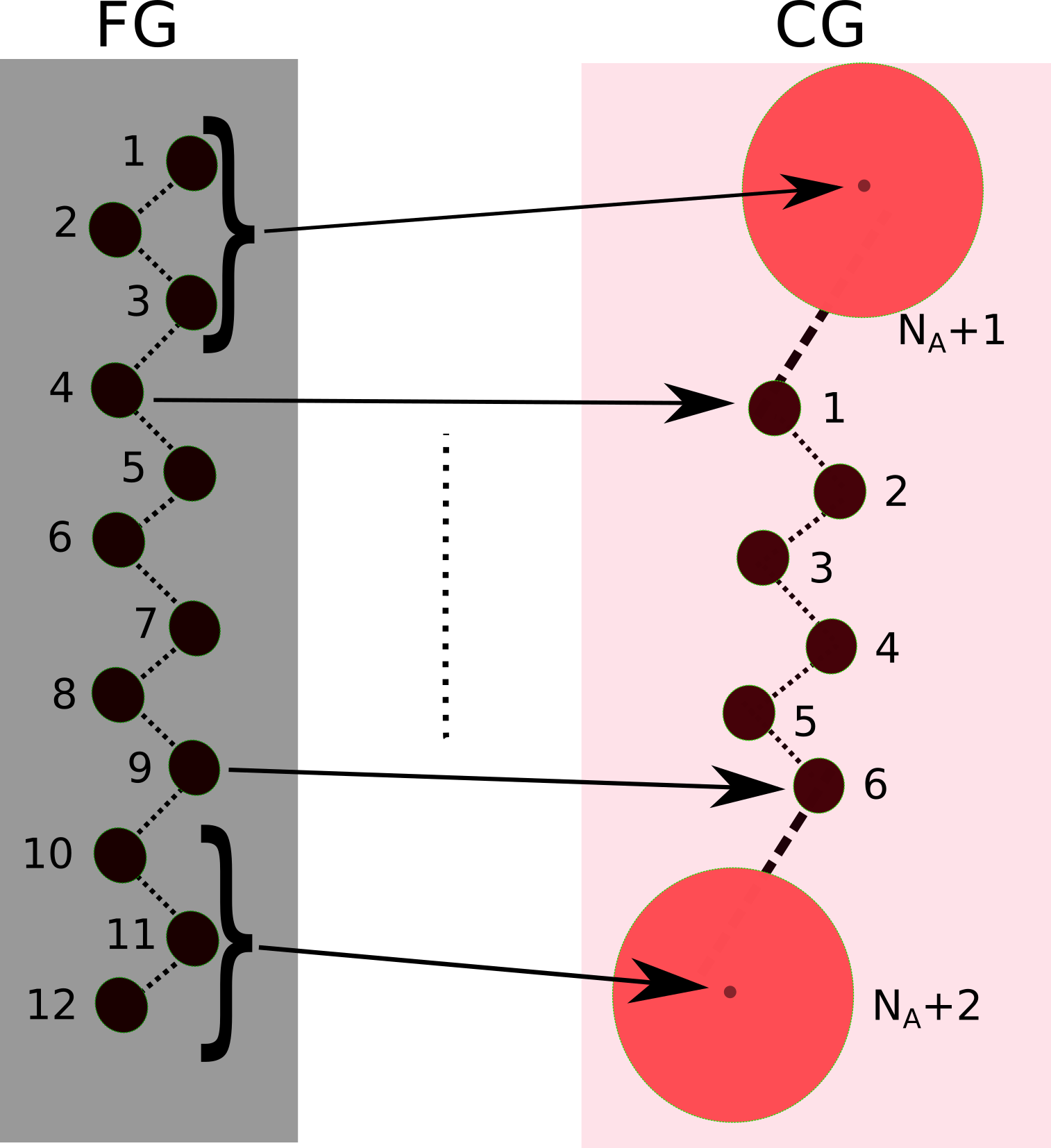}
    \caption{Sketch of the mapping of a molecule from the FG system to the CG system where the index mapping functions between the different resolutions are highlighted. In the picture, $\NA{}=6$, $\ncg=8$, therefore, $S_1=\{4\},\,S_2=\{5\},\,\ldots,\,S_6=\{9\},\,S_7=\{1,\,2,\,3\},\,S_8=\{10,\,11,\,12\}$}
        \label{Fig:sketch}
\end{figure}

Following \citep{Noid2008} we define $S_J$, the set of atoms included within the bead of index $J$:
\begin{equation*}
	S_J=\Big\{i \in \{1,\ldots,\nfg\}\,\; \mbox{such that atom}\,\;i\,\;\mbox{forms part of CG particle }J \Big\}.
\end{equation*}
Having assumed that each atom belongs to a single bead, the number of atoms corresponding to CG particle $J$ will be referred to as $s_J=\#S_J$. Given an atom with index $k$ in the FG system, we denote the index of the CG particle containing atom $k$ as $J_k$.
A sketch of a possible mapping between a molecule in FG system and the corresponding molecule in CG resolution is shown in \cref{Fig:sketch}.

A general CG (hybrid) system may include particles corresponding to single atoms (i.e. $J$ for which $s_J=1$), and into proper `beads' (i.e. $J$ for which $s_J>1$). The number of atoms in the CG system will be denoted $\NA{}:=\#\{J\text{ such that }s_J=1\}$, and the number of beads is denoted $N^B:=\ncg-\NA{}$. For notational convenience, we will assume that the indices are ordered in such a way that indices $J\in\{1,\ldots,\NA{}\}$ correspond to atoms, and $J\in\{\NA{}+1,\ldots,\ncg\}$ correspond to beads.
The above can be summarised in the following rules that apply throughout our derivation:
\begin{itemize}
\item Lower-case letters (e.g. $i,j$) refer to particles in the FG system.
\item Capital letters (e.g. $I,J$) label the particles (atoms or beads) in the CG system.
\item FG phase space variables are denoted with lower case letters: \\
  $$\Aphase{}=\cip{\Aphase{\bfr},\Aphase{\bfp}}=\cip{\bfr_1,\ldots,\bfr_{\nfg},\bfp_1,\ldots,\bfp_{\nfg}}.$$
\item CG phase space variables (also referred to as field or effective/averaged variables \citep{Kinjo2007,Guenza1999,Schweizer1989}) are denoted using capital letters:
  $$\Bphase{}=\cip{\Bphase{\bfR},\Bphase{\bfP}} = \cip{\bfR_1,\ldots,\bfR_{\ncg},\bfP_1,\ldots,\bfP_{\ncg}}.$$
\item The mass of the $j$th atom in FG system will be indicated with a lower case $m$ as $m_j$. The mass of the $J$th particle in CG system will be indicated as $M_J$.
\end{itemize}

\section{The Mori-Zwanzig formalism for Hamiltonian systems}\label{sec:zwanzig}

\noindent The Mori-Zwanzig (MZ) formalism is a reformulation of a dynamical system using projection operators. For application to MD, the dynamical system considered is Hamiltonian, and in this case there are two equivalent MZ approaches, akin to the Schr\"odinger and
Heisenberg `pictures' in the Copenhagen interpretation of Quantum Mechanics.
The former defines a projection operator acting on the Hilbert space of densities of states, and has been used in \citet{Kinjo2007}, resulting in a generalised Fokker-Planck equation for the CG variables. The latter considers a projection operator defined in the Hilbert space of all possible choices of CG variables, which may be considered observables of the FG system \citep{Nordholm1975}. This has been used in \citet{Hijon2009}, resulting in a Generalised Langevin Equation (GLE). The duality between spaces of observables and spaces of densities of states indicate the mathematical equivalence between these approaches. In what follows, we choose to follow the second approach, i.e. we write the evolution equations for the CG variables as functions of the underlying FG system, since this results in a more physically-intuitive derivation.

\subsection{The Mori-Zwanzig equations}
\noindent As stated in \cref{Sec:Def}, the state of the FG system is characterised by the variables $\Aphase{}$ in $6\nfg$-dimensional phase space. The evolution of the FG system is obtained by solving Hamilton's equations:
\begin{equation}\label{eq:hamiltonsyst}
	\totd{\Aphase{}(t)}{t} = \bfLambda\pard{\Ham(\Aphase{}(t))}{\Aphase{}}	
\end{equation}
with initial condition $\Aphase{}(0)=\Aphase{0}$, where $\Ham(\Aphase{}(t))$ is the Hamiltonian of the system and $\bfLambda$ is the $6\nfg\times 6\nfg$ symplectic matrix:
\[ \bfLambda =
\matrix{
    \bfzero & \bfI \\
    -\bfI & \bfzero \\
}.
\]
Here, $\bfI$ is the $3\nfg \times 3\nfg$ identity matrix and $\bfzero$ is a $3\nfg \times 3\nfg$ matrix of zeroes.

To derive a reduced description of this system, we consider a vector of \textit{observables}, $\Bphase{}(\Aphase{})$. Throughout the present section, these observables could be quite general,
but we will later apply the derivation made here to the particular choice $\Bphase{}(\Aphase{})=(\bfR_1(\Aphase{}),\ldots,\bfR_\ncg(\Aphase{}),\bfP_1(\Aphase{}),\ldots,\bfP_\ncg(\Aphase{}))$
corresponding to the positions and momenta of particles in the CG system. It is well-known \citep{Goldstein1980} that the evolution of such a
 function of phase variables is given by
\begin{equation}\label{Eq:liouville}
  \totd{\Bphase{}(\Aphase{}(t))}{t} =
  \Lio\Bphase{}(\Aphase{}(t)),\quad\Bphase{}(\Aphase{}(0)) = \Bphase{}(\Aphase{0})
  \qquad\text{where}\qquad
  \Lio := \sum_{i=1}^\nfg\cip{\pard{\Ham}{\bfp_i}\cdot\pard{}{\bfr_i} - \pard{\Ham}{\bfr_i}\cdot\pard{}{\bfp_i}}.
\end{equation}
The operator $\Lio$ is called the \emph{Liouvillean}, and in Poisson bracket notation, may be
written $\Lio = \{\Ham,\,\cdot\,\}$. We note that each of the terms in the definition of $\Lio$
is a dot-product. Using semigroup notation, \cref{Eq:liouville} has the formal solution:
\begin{equation}\label{Eq:solLio}
	\Bphase{}(\Aphase{}(t)) = \exp{t\Lio}\Bphase{}(\Aphase{0}).
\end{equation}
\Cref{Eq:solLio} is the exact dynamics of the observables. It is clear that this equation does not provide a self-consistent evolution equation for the observables, since in general, the evolution implicitly depends upon the full evolution of the FG system.

The Zwanzig projection \citep{Zwanzig1961,Nordholm1975} projects the equation for the observables $\Bphase{}$ onto the space of functions of the CG variables, which may be regarded as a subspace of the FG observable space.\footnote{This is true as long as some very broad measurability assumptions on the CG
  variables chosen are met; in practice, choosing CG variables which are continuous and differentiable in the FG variables is already sufficient, and this particular choice is implicitly assumed in the derivation presented here. In theory however, much more general choices can be made.}
The Zwanzig projection is equivalent to taking the conditional expectation $\aver{F}_{\Bphase{}}=\cexpval{F}{\,\Bphase{}}$ with respect to the equilibrium distribution, for a generic function $F=F(\Aphase{})$ acting on the phase space and for a specific value of the coarse-grained variables $\Bphase{}(\Aphase{})=\Bphase{}$ \citep{Chorin2000}. Since identifying the projection with a conditional expectation will be useful for the following sections, we provide a more detailed discussion in the SM. An alternative approach, the so-called Mori projection \citep{Mori1965}, can be seen as a finite-dimensional linear approximation of the Zwanzig projection \citep{Kauzlaric2011}; in certain cases it can be shown that the Mori projection and the Zwanzig projection are approximately equivalent \citep{Kauzlaric2011}.

We denote the Zwanzig projection as $\mathcal{P}_{\Bphase{}}$:
\begin{align}\label{Eq:proj}
	\mathcal{P}_{\Bphase{}}F(\Aphase{}) & = \aver{F}_{\Bphase{}}= \frac{1}{\Omega(\Bphase{})} \int{ F(\Aphase{}')}\,\dirac{\Bphase{}(\Aphase{}')-\Bphase{}}\rho^{eq}(\Aphase{}')\de\Aphase{}'
\end{align}
where the normalisation factor, also called the \textit{structure function} \citep{Zwanzig1961}, represents the number of micro-states compatible with the macro-state $\Bphase{}(\Aphase{})=\Bphase{}$:
\begin{equation}\label{eq:structurefunc}
	\Omega(\Bphase{}) = \int\dirac{\Bphase{}(\Aphase{}')-\Bphase{}}\mu(\de\Aphase{}').
\end{equation}
Here, $\mu(\de\Aphase{})=\rho^{eq}(\Aphase{})\de\Aphase{}$ is an equilibrium probability measure which must be specified. The conditional expectation just defined depends crucially on the choice of the underlying measure with respect to which expectations are taken. Different choices are possible and the modeller must jointly choose both the dynamics and the ensemble with respect to which they condition, as each ensemble will give slightly different forms of evolution equations for the coarse-grained system. The most common choices are usually either $\rho^{eq}(\Aphase{}) = \frac{1}{\partf}\exp{{-\beta \Ham(\Aphase{})}}$, i.e. the canonical ensemble, or $\rho^{eq}(\Aphase{}) = \delta(\Ham(\Aphase{})-E)$, i.e. the micro-canonical ensemble. The rest of the derivation will be obtained by using the canonical measure as probability equilibrium measure, but it is also perfectly possible to carry out the same derivation using the micro-canonical equilibrium measure.

Let $\qproj_{\Bphase{}}=\pproj_{\Bphase{}} - \idproj$ be the projection orthogonal to $\pproj_{\Bphase{}}$ with $\idproj$ being the identity operator. Splitting the identity by writing $\idproj=\pproj_{\Bphase{}}+\qproj_{\Bphase{}}$ on the right-hand side of \cref{Eq:liouville}, and performing some manipulations (for a more detailed derivation, we refer to the SM), a Generalised Langevin Equation (GLE) \citep{Hijon2009,Zwanzig1961,Darve2009} can be obtained:
\begin{align}
\label{eq:mz}
	\totd{\Bphase{}(t)}{t} = \Lio_{\pproj}\Bphase{}(t) &- \int_{0}^{t}{ \bffricmat(\Bphase{}(t-s),s)\pard{}{\Bphase{}}\entropy(\Bphase{}(t-s))\de s}\nonumber \\ &+ \beta^{-1}\int_{0}^{t}{\pard{}{\Bphase{}}\bffricmat\cip{\Bphase{}(t-s),s}\de s} + \fprojbf_{\Bphase{}}(t,\Aphase{0})
\end{align}
where  $\Lio_{\pproj}=\pproj_{\Bphase{}}\Lio$ is the projected Liouvillean, the application of which yields functions depending only on $\Bphase{}$. 
Recalling that the underlying equilibrium measure is the canonical ensemble, the function $\entropy$ is
defined to be
\begin{equation}\label{Eq:FreeEnergy}
  \entropy = -\frac{1}{\beta}\ln{\Omega(\Bphase{})},
\end{equation}
and, since it has units of energy, $\entropy$ is interpreted as a contribution to the free energy of the system, in view of the definition of $\Omega(\Bphase{})$ in \cref{eq:structurefunc}.\footnote{In the micro-canonical ensemble, i.e. with microcanonical equilibrium measure, the free energy instead becomes
$\mathcal{S}(\Bphase{})=-E\log \Omega(\Bphase{})$.}
We note that in \citep{Hijon2009}, the alternative quantity $\mathcal{S}(\Bphase{}) = \kb\ln{\Omega(\Bphase{})}$ is defined, which is an entropy. Both definitions are identical up to a multiplicative factor, and depend upon on the assumed temperature of the system.

The last term $\fprojbf_{\Bphase{}}(t,\Aphase{0})$ is defined as:
\begin{equation}\label{eq:randomforce}
	\fprojbf_{\Bphase{}}(t,\Aphase{0}) = \exp{{t\Lio_{\qproj}}}\Lio_{\qproj}\Bphase{}(\Aphase{0})
\end{equation}
where $\Lio_{\qproj}=\qproj_{\Bphase{}}\Lio$ is the projected dynamics operator that still depends on the FG variables $\Aphase{}$. The quantity $\fprojbf_{\Bphase{}}(t,\Aphase{0})$ is given by the solution of an auxiliary set of equations called the orthogonal dynamics equations \citep{Givon2004,Givon2005}, which we briefly discuss in the following sections. It is important to highlight here that the complexity of the \cref{eq:mz} is not reduced with respect the starting point (\cref{Eq:liouville}), it has simply been shifted into the calculation of the orthogonal dynamics. However, it may be argued that close to equilibrium, this term is uncorrelated with $\Bphase{}$ and therefore approximated as a random noise \citep{Evans1990} (see SM).

The friction matrix $\bffricmat$ is a positive definite matrix whose components can be written in Green-Kubo form. In the case where the underlying measure is the canonical ensemble, it is defined to be:\footnote{In the microcanonical ensemble this definition becomes $\bffricmat(\Bphase{},t) = \frac{1}{E}\proj{\Bphase{}}(\fprojbf_{\Bphase{}}(0,\cdot)\otimes\fprojbf_{\Bphase{}}(t,\cdot))$.}
\begin{equation}\label{Eq:fricmatrix}
	\mathcal{M}(\Bphase{},t) = \beta\,\proj{\Bphase{}}\cip{\sqp{\Lio_{\qproj} \Bphase{}}\otimes\sqp{\exp{t\Lio_{\qproj}}\Lio_{\qproj}\Bphase{}}} = \beta\,\proj{\Bphase{}}\Big(\fprojbf_{\Bphase{}}(0,\cdot)\otimes\fprojbf_{\Bphase{}}(t,\cdot)\Big).
\end{equation}
where $\otimes$ represents a tensor product.

The terms involving the friction matrix turn out to be the most difficult to compute in a real calculation because of the memory effect. In this work we will focus on the practical derivation of the relevant equations for a CG atomic system including both atoms and beads. We will briefly discuss the friction matrix and the memory effects but a more thorough analysis of these terms is left for future publications.
A detailed derivation of \cref{eq:mz} and \cref{Eq:fricmatrix}, following \citet{Hijon2009}, is reported in the SM. We remark that this derivation is quite general, requiring solely that $\Bphase{}$ is a sufficiently smooth function of $\Aphase{}$.

In \cref{sec:analysis} we will derive more explicit forms of the terms appearing in \cref{eq:mz} for a CG system,  for specific choices of the CG variables which arises naturally from physical considerations, namely canonical mappings (i.e. transformations).
We will show in \cref{Sec:Res} how these formal expressions can be computed from MD simulations for a simple system represented by a one dimensional chain composed by atoms with different mass connected through a Lennard-Jones potential. The scale separation hypothesis will be discussed for this simple system.

\section{Canonical mappings and coarse-graining}
\label{sec:analysis}

In the previous section, the CG variables considered were quite general. We now consider a
more concrete set of choices defining a coarse-graining procedure for an atomistic system.
We suppose that the Hamiltonian for the FG system takes the form
\begin{equation}\label{Eq:Hamiltonian}
	\Ham(\bfr,\bfp) = {\sum_{i=1}^{\nfg}{\frac{\bfp_{i}^2}{m_{i}} }} + \pot(\bfr),
\end{equation}
where $\pot(\bfr)$ is the interaction potential between atoms and $m_i$ is the mass of the $i$th atom. As is usual for molecular models, $U$ is typically written as a sum of $2$, $3$ and
$4$-body potentials, chosen to accurately capture interatomic interactions.

\subsection{A particular class of mapping operators}\label{sec:mappop}

\noindent In the terminology of \cref{Sec:Def}, the mapping operators which fix the CG variables will be observables of the FG system. These are functions $\Psi,\Phi:\real^{6\nfg}\to\real^{6\ncg}$ such that
\begin{align}%
	\Bphase{\bfR}  & = \Psi (\Aphase{}) \nonumber \quad\text{and}\quad
	\Bphase{\bfP}  = \Phi (\Aphase{}).
\end{align}
The simplest choice for a CG mapping is to simply select the position of a representative atom from each bead, along with its conjugate momentum. This particular mapping is used in CG models where the bead is centred on one of the atoms belonging to the bead itself (e.g. see the coarse-grained model for poly-ethylene presented in \citep{DiPasquale2012}):
\begin{alignat}{3}
	\bfR_J=\Psi_J (\Aphase{}) & = \bfr_i,&\quad&\text{for some }i \in S_J, \label{Eq:constr3} \\
    \bfP_J=\Phi_J (\Aphase{}) & = \bfp_i ,&\quad&\text{for some }i \in S_J.  \label{Eq:constr4}
  \end{alignat}
  With this definition, the construction of the projection operator is particular straightforward: we need simply to `integrate out' all other positions and momenta.

The class of mapping operators can be extended if, before selecting a single degree of freedom for each CG particle, we first perform a \emph{canonical transformation}.
Canonical transformations have the attractive property that they preserve the Hamiltonian structure of the model.\footnote{In the language used in the mathematical literature, canonical transformations belong to the wider class of `symplectomorphisms'.} If the pair of functions $(\bfzeta,\bftheta):\real^{6\nfg}\to\real^{6\nfg}$
constitute a change of variable on the FG phase space, a simple criterion for the mapping to be a canonical transformation is that it preserves the Poisson bracket, which requires that
$$
\sum_{i=1}^\nfg\frac{\partial \bfzeta_k(\Aphase{})}{\partial\bfr_{i}}\cdot\frac{\partial\bftheta_l(\Aphase{})}{\partial \bfp_{i}} = \delta_{kl}\quad\text{and}\quad
\sum_{i=1}^\nfg\frac{\partial \bfzeta_k(\Aphase{})}{\partial\bfp_{i}}\cdot\frac{\partial\bftheta_l(\Aphase{})}{\partial \bfr_{i}} = -\delta_{kl}
$$
where $\delta_{kl}$ is the Kronecker delta.
In general, we will define a class of \emph{canonical coarse-graining operators}, being the composition of the index selection operator with a canonical transformation, i.e. for each $J=1,\ldots,\nfg$, setting $\Psi_J = \bfzeta_i$ and $\Phi_J = \bftheta_i$ for some $i$, depending on $J$. If the canonical transformation involved is linear, then the entire CG mapping operator is linear, and we will call mappings in this class the \emph{linear canonical coarse-graining operators}.

One of the most often considered choices within this class is to define the mapping operator to select the centre of mass of a CG particle as a positional coordinate, and the relevant conjugate momentum, which is simply the sum of momenta of all atoms within the bead, i.e.
\begin{align}
	\bfR_{I}=\Psi_{I} (\Aphase{}) & = \frac{1}{M_I}\sum_{k\in S_I} m_k\bfr_k = \sum_{k\in S_I} c_k\bfr_k \label{Eq:constr1} \\
     \bfP_{I}=\Phi_{I} (\Aphase{}) & = \sum_{k\in S_I}{\bfp_k}. \label{Eq:constr2}
\end{align}
where $M_I=\sum_{k\in S_I}m_i$ and for each $k$, $c_k=m_k/M_I>0$ where $I$ is such that $k\in S_I$.
This is a linear canonical transformation, and has the convenient feature that it is local: each transformed degree of freedom depends only on a small number of FG variables, all of which are close to one another in space. For a hybrid description, the CG variables which correspond to single
atoms are unchanged, since $c_k=1$ for $k$ for which $S_I=\{k\}$.

The particular linear canonical mapping operators defined above are convenient due to their
immediate physical meaning; however, it is legitimate to ask if there are other ways in which to better characterise the behaviour of beads, particularly in view of the further practical approximations we will make below. In some cases, the centre of mass of a group of atoms could preserve too little information about the system: instead, the choice of the centre of mass could be complemented with additional degrees of freedom which are nonlinear functions of the atomic positions within a bead, in order to capture rotational degrees of freedom when beads are expected to have an anisotropic structure. The price to pay for the additional detail is in more complex implementation and less intuition about the resulting
equations; nevertheless, this is a direction we aim to explore further in subsequent work.

The linear canonical coarse-graining operators defined in \tcref{Eq:constr1,Eq:constr2} are not injective functions of the fine-grained phase space. This property means that the same centre of mass and bead momentum may be given by different combinations of atomistic positions and momenta. This can be easily shown by considering a two atoms $\{\bfr_1,\bfr_2\}$ bead, with masses $\{m_1, m_2\}$ and $m_1\neq m_2$. The position of the centre of mass will be on the line connecting the two atoms. If the two atoms are swapped, the centre of mass will be on the same exact position, i.e. $\Psi(\bfr_1,\bfr_2)=\Psi(\bfr_2,\bfr_1)$. In vacuum the swapping will not have any effect, but in a system with finite density (i.e. in a system where atoms $1$ and $2$ are surrounded by other atoms) the two physical configurations will be different because, in general, the energies of the two systems will be different. The non injectivity of the mappings in \tcref{Eq:constr1,Eq:constr2} reflects the loss of information due to the reduction of the degrees of freedom in the coarse-graining procedure\footnote{In the context of the relative entropy framework \tcitep{ScottShell2008}, which will be further discussed in a later section (see \tcref{Sec:effPot}), the non injectivity of the mapping operator is the reason why the mapping entropy term arises (see Eq. 4, \tcitep{ScottShell2008}), as a measure of the degeneracy of the mapping. Higher mapping entropy values correspond to more atomistic configurations that can be mapped into the same CG variables. }.

Related to this loss of information is the so-called \emph{back-mapping problem} \tcitep{Peter2009,Peter2010}, which is the problem of obtaining an inverse of the position mapping operator (\tcref{Eq:constr1}), given some additional constraints, which can be either geometrical \tcitep{Santangelo2007} or may involve energy minimisation \tcitep{Chen2009}. We will give a brief discussion about the use of Mori-Zwanzig projection operator in the back-mapping problem when we introduce the orthogonal dynamics (see \tcref{Sec:OrthD}).

\subsection{Separability and effective interactions}\label{Sec:EffInt}

\noindent Typically the Hamiltonian of a molecular system is expressed additively as sum of the kinetic energy and a potential energy expressed as a sum of 2, 3, and 4-body interactions. When considering a hybrid CG system, this additive decomposition allows us to rewrite $\pot$ in the form
\begin{equation}\label{Eq:int}
	\pot(\bfr) = \pot^{AA}(\bfr) + \pot^{AB}(\bfr) + \pot^{BB}(\bfr)
\end{equation}
where $\pot^{AA}(\bfr)$ is the sum of interaction potentials between the particles in the CG system that are simply atoms, $\pot^{BB}(\bfr)$ is the sum of interactions between atoms that are grouped into beads in the CG system, and $\pot^{AB}(\bfr)$ is the sum of all other interactions, i.e. interactions involving both atoms which are part of beads and atoms which are not coarse-grained.

Following \citet{Espanol2009}, it is convenient to start by writing down explicitly the free energy of a generic atomistic/coarse-grained system. When considering a separable Hamiltonian and a CG
mapping of the form given in \cref{Eq:constr1,Eq:constr2}, the integral can be factorised into
contributions arising from the momenta of CG particles and their positions, which results in
the decomposition
\begin{align}
  \entropy(\Bphase{}) =-\frac{1}{\beta}\log{}\biggl(
  &\int{\frac{1}{\mathcal{Z}} \exp{{-\beta \Ham(\Aphase{})}}} 
    \dirac{\Psi\cip{\Aphase{}}-\Bphase{\bfR}}\dirac{\Phi\cip{\Aphase{}}-\Bphase{\bfP}}\de\Aphase{}
    \biggr) =\entropy_{\bfP}(\Bphase{\bfP})+ \entropy_{\bfR}(\Bphase{\bfR}).   \nonumber 
\end{align}
Computing, we find the two terms in the latter expression then correspond to:

\paragraph{the Kinetic Energy of the CG particles:}
		\begin{align}\label{Eq:MomB}
			\entropy_{\bfP}(\Bphase{\bfP})  
                  &=-\frac{1}{\beta} \log{}  \cip{\frac{1}{\mathcal{Z}_{\bfP}}\int{\de\Aphase{\bfp}' \mathrm{exp}\bigg(-\beta\sum_{i=1}^\nfg\frac{|\bfp_{k}|^2}{2m_k}}\bigg){\prod_{J=1}^{\ncg}{\dirac{\Phi_{J}(\Aphase{\bfp}')-\bfP_{J}} }}} \nonumber\\
				& = {\sum_{J=1}^{\ncg}{\frac{|\bfP_{J}|^2}{2M_{J}}}} + const,
		\end{align}

\paragraph{the Potential Energy of the CG particles:}
\begin{align}\label{Eq:EntrInt}
  \entropy_{\bfR}(\Bphase{\bfR})
  &=-\frac{1}{\beta} \log{} \bigg(\int\de\Aphase{\bfr}' \frac{1}{\mathcal{\mathcal{Z}_{\bfR}}}
    \exp{-\beta \big[\pot^{AA}(\bfr')+\pot^{AB}(\bfr)+\pot^{BB}(\bfr')\big]}\prod_{J=1}^{\ncg}
    \dirac{\Psi_{J}(\bfr')-\bfR_{J}} \bigg) \nonumber 	\\
  & = \pot^{AA}\Big(\{\bfR_I\}_{I=1}^{\NA{}}\Big)  \nonumber\\
  &\qquad - \frac{1}{\beta} \log{}  \cip{\int{\de\Aphase{\bfr}' \frac{1}{\mathcal{\mathcal{Z}_{\bfR}}}
    \exp{\big[{{-\beta \big(\pot^{AB}(\{\bfR_I\}_{I=1}^{\NA{}},\bfr') +\pot^{BB}(\bfr') \big)}\big]}}}
    \prod_{J=\NA{}+1}^{\NB{}}\dirac{\Psi_{J}(\bfr')-\bfR_{J}}}  \nonumber 	\\
  & = \pot^{AA}\Big(\{\bfR_I\}_{I=1}^{\NA{}}\Big) + \Veff(\Bphase{\bfR}).
\end{align}
In the latter derivation, we have used the fact that the $\delta$ function over the atoms can be trivially applied. The integral in the second line must be intended as performed over all the coordinates of atoms belonging to some bead. The integral in the second line is also defined to be the latter term on the third line, $\Veff$.
$\Veff$ can be interpreted as an effective potential due to the property that if $I$ satisfies
$S_I=\{k\}$, then
\begin{equation}\label{Eq:ForceAt}
	-\pard{\Veff}{\bfR_I} = \aver{-\pard{\pot^{AB}}{\bfr_k}}_{\Bphase{}} =  \aver{\bfF_k}_{\Bphase{}}
\end{equation}
where $\bfF_k$ is the force acting on atom $k$ in FG system, which corresponds to particle $I$ in
the CG system. An explicit derivation of this property involves the rewriting of the integral above as a Mori-Zwanzig projection, as shown in the SM.  A similar result can be obtained for beads:
\begin{align}\label{Eq:VeffB}
	-\pard{\Veff}{\bfR_{J}} 	 = \aver{-\sum_{k\in S_J}\frac{1}{c_ks_J}\pard{}{\bfr_k}\cip{\pot^{AB}+\pot^{BB}}}_{\Bphase{}}  = \aver{\Phi_J(\bfF)}_{\Bphase{}}
\end{align}
where $\Phi_J(\bfF)$ is the sum of the forces acting on the atoms in bead $J$.

It is important to remember that this definition of $\Veff$ is still formal and, in general, cannot be computed analytically as it depends on all coarse-grained variables. To make such a definition useful in practice, further approximation and assumptions will be needed, as discussed in \cref{sec:discussions}.

Following \tCitet{Espanol2009}, we note that \tcref{Eq:VeffB} represents the average force induced by all the atoms in the system not included in the bead $J$ on the CG coordinate $\bfR_J$. The former interpretation is equivalent to the one used in the Force Matching method \tcitep{Noid2008} which derives the CG interactions from the forces calculated in an atomistic simulation.
In \tCitet{Noid2008} the effective potential was obtained by invoking the consistency condition, which asks that the equilibrium probability density of the CG system for a certain choice of the mapping operator is equal to the equilibrium probability density of the atomistic system constrained by the mapping operator (see Eq. 20 in \tcitep{Noid2008}). In our case we showed that the consistency condition is automatically included in the MZ projection operator and solving the problem for the determination of the CG potential with Force Matching means finding a suitable approximation of the $\Veff$ potential. Further discussion of possible approximations of the effective potential is provided at the end of \tcref{Sec:effPot}.

\subsection{Projected dynamics}\label{sec:projcoor}

\noindent To explicitly write \cref{eq:mz} for the CG variables we need to first compute the dynamics of $\Bphase{}(\Aphase{})$. For the position and momentum of a particle $\{\bfR_{J},\bfP_{J}\}$ in the CG system:
\begin{align}
	\Lio\bfR_{J} & = \sum_{k\in S_J}{\frac{\bfp_k}{m_k}\pard{\Psi_J(\Aphase{\bfr})}{\bfr_k}}  = \sum_{k\in S_J}{\frac{\bfp_k}{m_k}\frac{m_k}{M_J}}  = \frac{\bfP_J }{M_{J}}  \label{Eq:LPB}  \\
	\Lio\bfP_{J} & = -\sum_{k\in S_J}{\pard{\pot(\Aphase{\bfr})}{\bfr_k}}\pard{\Phi_J(\Aphase{\bfp})}{\bfp_k}  = -\sum_{k\in S_J}{\pard{\pot(\Aphase{\bfr})}{\bfr_k}}
\end{align}
where we have simply used the definition of $\Phi,\Psi$ as chosen in \cref{Eq:constr1,Eq:constr2}.
We note that by making this particular choice, we obtain that $\Lio\bfR_{J}$ is written in terms of the respective CG momentum, and therefore $\Lio_\pproj \bfR_J=\frac{1}{M_J}\bfP_J$. For the same reason, if $\qproj_{\Bphase{}}$  is applied to \cref{Eq:LPB}, the result vanishes, i.e. $\Lio_\qproj\bfR_{I}=0$.

Considering the CG momenta, we have
\begin{equation*}
  \begin{aligned}
    \Lio_\pproj\bfP_{I} & = -\pard{\pot^{AA}}{\bfR_I} - \pard{\Veff}{\bfR_{I}}	&\quad&\text{for }I=1,\ldots,\NA{},\text{ with }S_I=\{k\},&\\
    \Lio_\pproj\bfP_{J} & = - \pard{\Veff}{\bfR_{J}}	 &\quad&\text{for }I=\NA{}+1,\ldots,\ncg,&
  \end{aligned}
\end{equation*}
and therefore the orthogonal projections are
\begin{equation}\label{eq:orthForces}
\begin{aligned}
  \Lio_\qproj\bfP_{I} & =  -\pard{\pot^{AB}}{\bfr_k}  + \pard{\Veff}{\bfR_I} &\quad&\text{for }I=1,\ldots,\NA{},\text{ with }S_I=\{k\},& \\
  \Lio_\qproj\bfP_{J} & = -\sum_{k\in S_J}{\pard{}{\bfr_k}\cip{\pot^{AB}+\pot^{BB}} } + \pard{\Veff}{\bfR_J}	&\quad&\text{for }I=\NA{}+1,\ldots,\ncg.&
\end{aligned}
\end{equation}
These functions are often referred to as the \emph{fluctuating forces}, and we therefore define for a generic time $t$:
\begin{align}
	\fprojbf_I(t,\cdot) & = \exp{t\Lio_\qproj}\Lio_\qproj\bfP_I(\cdot) &\quad&\text{for }I=1,\ldots,\ncg.& \label{Eq:StocFor}
\end{align}

\subsection{Invariants and properties of the Friction Matrix}

\noindent The Friction Matrix, defined in \cref{Eq:fricmatrix}, represents the non-Markovian contribution to
the dynamics of the CG variables: the fact that the system `remembers' earlier states can be seen
from the time convolution with all the previous states along the trajectory. Nevertheless, we can
deduce some important properties of $\fricmat$, which will inform the choices we make in
\cref{sec:discussions} to approximate it.

As observed in \cref{sec:projcoor}, when choosing a linear canonical CG operator, $\Lio_\qproj\bfR=0$. It follows that the only non-zero terms in the friction matrix arise from the correlations between fluctuating forces $\fprojbf$, defined in \cref{Eq:StocFor}. Furthermore, since $\Lio_\qproj\bfP_I$ is expressed in terms of a function of $\bfr$ alone (see \cref{eq:orthForces}), it follows that $\fricmat$ must depend only on $\bfR$, and not on $\bfP$.

By using the conservation of linear momentum, which states that $\Lio\cip{{\sum_{J=1}^{\ncg}{\bfP_J}}}=0$, we see that $\Lio_\qproj\cip{{\sum_{J=1}^{\ncg}{\bfP_J}}}=0$, and hence 
\begin{align}
	\exp{t\Lio_\qproj}\Lio_\qproj\sum_{J=1}^{\ncg}\bfP_{J} = 0.
\end{align}
Using this observation and the definition of $\fricmat$ (\cref{Eq:fricmatrix}), we find that
\begin{align} \label{Eq:fricmatGalI}
	\fricmat_{II}  & = \beta\,\proj{\Bphase{}}\cip{\sqp{\Lio_{\qproj} \bfP_I}\sqp{\exp{t\Lio_{\qproj}}\Lio_{\qproj}\bfP_I}} =-\beta\,\proj{\Bphase{}}\bigg(\Lio_{\qproj} \bfP_I\,\exp{t\Lio_{\qproj}}\Lio_{\qproj}\sum_{K\neq I}\bfP_K\bigg) =- \sum_{K\neq I}{\fricmat_{IK}}.
\end{align}

\subsection{Final system of equations}

\noindent Now that every term of \cref{eq:mz} is explicitly written, the evolution for the CG system is given by solving the following system of equations:
\begin{align}
  \pard{\bfR_{K}}{t}
  &= \frac{\bfP_{K}}{m_{K}} \nonumber \\
  \pard{\bfP_{K}}{t}
  & = -\omega_{\NA{} K}\pard{\pot^{AA}}{\bfR_{K}}-\pard{\Veff}{\bfR_{K}} - \sum_{J=1}^{\ncg}{\int_{0}^{t}{\fricmat_{KJ}}(t-s,s)\frac{1}{M_{J}}\bfP_{J}(t-s) }\,\de s \nonumber \\
  & + \beta^{-1} \sum_{j=1}^{\ncg}{\int_{0}^{t}{\de s \pard{}{\bfR_{J}}\fricmat_{KJ}(s)}}
    + \fprojbf_{K}(t)\label{Eq:DevB} 
\end{align}
where $\omega_{N^AK}$ is equal to zero whenever the difference between the first and second subscript is negative and one otherwise.


\subsection{Orthogonal Dynamics}\label{Sec:OrthD}

\noindent The last term of \cref{Eq:DevB}, identified in first approximation as noise terms, is function of the evolution of the orthogonal components of the projection, i.e. all the terms where the orthogonal projection $\qproj_{\Bphase{}}$ is present. In general, the noise function $v(\Aphase{},t) = e^{t\Lio_{\qproj}}v(\Aphase{},0)$, is defined as the solution of the orthogonal system \citep{Givon2004,Givon2005}: 
\begin{align}\label{eq:orthdyn}
	\pard{}{t}v(\Aphase{},t) & = \Lio_{\qproj} v(\Aphase{},t) \nonumber \\
	v(\Aphase{},0) & = \qproj_{\Bphase{}} \alpha(\Aphase{})
\end{align}
where $\alpha$ is a generic function of the FG phase-space $\Aphase{}$, and with $\pproj_{\Bphase{}} \qproj_{\Bphase{}} \alpha(\Aphase{}) = 0$, i.e. $\qproj_{\Bphase{}} \alpha(\Aphase{})$ is a function in the null space of $\pproj$. The existence of the solutions for the orthogonal dynamics equations was discussed by \citet{Givon2005}. Since the application of $\pproj_{\Bphase{}}$ and differentiation with respect to time
commute, it follows that $v$ remains in the null space of $\pproj_{\Bphase{}}$ for all time.
In the particular case of the CG mapping described in \cref{sec:mappop}, we find that $v_{\bfR}(\Aphase{},0)=0$, whereas for momenta, $v_{\bfP}(\Aphase{},0)=\fprojbf_I(0)=\Lio_{\qproj}\Bphase{\bfP}$. By using the fact that trajectories generated by orthogonal dynamics must be in the null space of $\pproj$ at all times, we can describe the orthogonal dynamics by using constrained Hamiltonian dynamics, where the constraints are represented by the mapping functions for positions and momenta defined earlier (\cref{Eq:constr1,Eq:constr2}). In particular, the mapping defined in \cref{Eq:constr1,Eq:constr2} permits us to write the orthogonal equations explicitly as constrained equations \citep{Akkermans2000}. The equivalence of orthogonal and constrained dynamics is shown by rewriting the FG Hamiltonian with centre of mass and internal coordinates (similar to the computation shown in SM eq. S19). The orthogonal projection operator in this case discards the centre of mass coordinates, and the operator $\Lio_{\qproj}$ describes the evolution of the orthogonal coordinates, which can be chosen to
correspond to internal coordinates in each bead. In the framework of the constrained dynamics, if we write the constrained Lagrangian for this system, we can derive a Hamiltonian, and a Liouville operator for the constrained system equivalent to $\Lio_{\qproj}$ \citep{Akkermans2000}.
The constraints, \cref{Eq:constr1,Eq:constr2}, can be written as:
\begin{align*}
	\Psi_I(\bfr) - \bfR_I & = 0 \\
	\Phi_I(\bfp) - \bfP_I & = 0
\end{align*}
The equation of motion for the constrained dynamics are given by:
\begin{equation}
	\totd{\bfr_k}{t} = \bfp_k - \frac{\bfP_I}{m_{I}},\qquad
	\totd{\bfp_k}{t} = \pard{U(\bfr)}{\bfr_k} - \frac{m_k}{M_I}\bfF_I\label{Eq:orth} 
\end{equation}
where $I$ is the index such that $k\in S_I$, and $\bfF_I$ is given by the expressions in \cref{eq:orthForces}. The derivation for previous equations is reported in the SM. Simulating them provides a tractable approach to computing the fluctuating forces, and therefore to describe the memory kernel through the Green-Kubo relation given in \cref{Eq:fricmatrix}: indeed, this is exactly the approach we take in the numerical example presented in \cref{Sec:Res}.

In \tcref{sec:mappop} we gave a brief introduction to the back-mapping problem, which involves finding atomistic positions which are consistent with CG variables. While a comprehensive discussion of this topic is beyond the scope of this paper, we note that the orthogonal dynamics presented here provide a way to sample the space of configurations which are consistent with the coarse-grained variables. That means that given a CG configuration, solutions of the back-mapping problem can be obtained by evolving the system under equations \tcref{Eq:orth}, since solutions will automatically fulfil all the constraints (i.e. bonds, atoms, dihedrals), preserving the CG variables. This approach can naturally explore the whole (compatible) phase-space and find its probability density, clearly showing the intrinsic uncertainty in the back-mapping problem, instead of arbitrarily introducing ad-hoc minimisation of the constraints or the energy.

\section{Discussion}\label{sec:discussions}

\noindent  In this section, we discuss the relationship between the derivation and modelling choices presented above, and other approaches in the literature, as well as the Markovian approximation, an approximation which renders the MZ equations \cref{Eq:DevB} more tractable in practice.

\subsection{Approaches to the Effective Potential}\label{Sec:effPot}

\noindent The Mori-Zwanzig projection of the equations of motion for coarse-grained variables leads to the definition of the effective potential $\Veff$ in \cref{Eq:VeffB}. This is, in general, a complex function on a high-dimensional space since it depends on all coarse-grained variables, and is therefore difficult to compute. 
In \cref{Sec:ApproxVeff} we will discuss about approximations often (sometimes implicitly) used to simplify it and make it practically computable. While it is not our intention to propose a new technique to compute the effective potential, here we wish to highlight the fact that the effective potential derived in the MZ framework, and those used in other CG approaches are all closely related.

The effective potential \cref{Eq:VeffB} can be connected with the widely used concept of the Potential of Mean Force (PMF) \citep{Kirkwood1935}, $\pfm{n}$, which represents the potential that returns the average force acting on an atom $j$ integrated over all the configurations of the atoms $n+1,\ldots,M^{FG}$ while atoms $1,\ldots,n$ are kept fixed \citep{Chandler1987}:
\begin{equation}\label{Eq:pfm}
	-\pard{\pfm{n}}{\bfr_j}  = \frac{\int{\de \bfr'_{n+1}\cdots \de \bfr'_{M^{FG}} \cip{-\pard{U(\Aphase{\bfr})}{\bfr_j} } e^{-\beta U(\bfr)}}}{\int{\de \bfr'_{n+1}\cdots \de \bfr'_{M^{FG}} e^{-\beta U(\bfr)}}}
	=
	\frac{\int{\de \Aphase{\bfr}' \cip{-\pard{U(\Aphase{\bfr}')}{\bfr_j} } e^{-\beta U(\Aphase{\bfr}')}  \prod_{i=1}^{n}\dirac{\bfr_i' - \bfr_i} }  }{\int{\de \Aphase{\bfr}'  e^{-\beta U(\Aphase{\bfr}')}\prod_{i=1}^{n}\dirac{\bfr_i' - \bfr_i} }}.
\end{equation}
 This can be easily identified with the effective potential defined in \cref{Eq:VeffB} when using the mapping functions defined in \cref{Eq:constr3,Eq:constr4} (i.e., CG positions identified by selected atoms):
\begin{equation}\label{Eq:MZpfm}
-\pard{\Veff}{\bfr_j} = \frac{\int{\de \Aphase{\bfr}' \cip{-\pard{U(\Aphase{\bfr}')}{\bfr_j} } e^{-\beta U(\Aphase{\bfr}')}  \prod_{i=1}^{n}\dirac{\bfr_i' - \bfr_i} }  }{\int{\de \Aphase{\bfr}'  e^{-\beta U(\Aphase{\bfr}')}\prod_{i=1}^{n}\dirac{\bfr_i' - \bfr_i} }}=-\pard{\pfm{n}}{\bfr_j}
\end{equation}
However, this equivalence is only valid for a particular choice of the mapping function in CG system. We can therefore define
a \textit{Generalised Potential of Mean Force}, $W^{(\ncg)}$ (GPFM), averaging forces on atoms, while using a more general arbitrary mapping function as a constraint.
If we consider the force acting on the atom $J_k$ averaged over the configuration of all the other atoms, keeping fixed the centres of mass of $\ncg$ beads, we will obtain the following expression for the GPFM
\begin{equation}\label{eq:gpmf}
	-\at{{\pard{W^{(\ncg)}}{\bfr_{J_k}}}}{\bfR_1,\ldots,\bfR_{\ncg}}
	=
	\frac{\int{\de \Aphase{\bfr} \cip{-\pard{U(\bfr)}{\bfr_{J_k}} } e^{-\beta U(\bfr)}  \prod_{I=1}^{\ncg}\dirac{\Psi_I - \bfR_{I}} }  }{\int{\de \Aphase{\bfr}  e^{-\beta U(\bfr)}\prod_{I=1}^{n}\dirac{\Psi_I - \bfR_{I}} }},
\end{equation}
where $J_k$ maps the general index in the FG space to the index in the CG space. By using the same symbols defined in \cref{Sec:Def} for the number of atoms in the group $J$, $s_J$, we divide both side of \cref{eq:gpmf} by $s_Jc_{J_k}$, where $c_{J_k}$ is defined in \cref{Eq:constr1}, and summing over $s_J$ we obtain the relation between the GPMF and the effective potential:
\begin{equation}\label{Eq:gpmf}
	\pard{W^{(\ncg)}}{\bfR_J} =
	\sum_{k=1}^{s_J}\frac{1}{s_Jc_{J_k}}\pard{W^{(\ncg)}}{\bfr_{J_k}} =-
	\frac{\int{\de \Aphase{\bfr} \sum_{k=1}^{s_J}\cip{-\frac{1}{s_Jc_{J_k}}\pard{U(\hr)}{\bfr_{J_k}} } e^{-\beta U(\bfr)}  \prod_{I=1}^{\ncg}\dirac{\Psi_I - \bfR_{I}} }  }{\int{\de \Aphase{\bfr}  e^{-\beta U(\bfr)}\prod_{I=1}^{n}\dirac{\Psi_I - \bfR_{I}} }}
=	\pard{\Veff}{\bfR_J}
\end{equation}
where we have used the fact that $\pard{\bfR}{\bfr}=\pard{\Psi}{\bfr}=c$, and, with a slight abuse of notation, $W$ can be therefore thought as a function of $\bfr$ (through the map $\Psi$) or $\bfR$.
The definition of GPMF given above can be easily extended to the case where only a subset, $\{1,\ldots,n\}$, say, of the $\ncg$ centres of mass are fixed, and a further averaging is performed over all configurations of the remaining $\{n+1,\ldots,\ncg\}$ centres of mass. In this case we obtain:
\begin{equation}\label{eq:extgpmf}
	-\at{{\pard{W^{(n)}}{\bfr_{J_k}}}}{\bfR_1,\ldots,\bfR_{n}} = \frac{\int\de\bfR_{n+1}\cdots\de \bfR_{\ncg}\int{\de \Aphase{\bfr}\cip{-\pard{U(\hr)}{\bfr_{J_k}} } e^{-\beta U(\bfr)}  \prod_{I=1}^{\ncg}\dirac{\Psi_I - \bfR_{I}} }  }{\int\de\bfR_{n+1}\cdots\de \bfR_{\ncg}\int{\de \Aphase{\bfr}  e^{-\beta U(\bfr)}\prod_{I=1}^{n}\dirac{\Psi_I - \bfR_{I}} }}
\end{equation}
It is important to highlight here that \cref{eq:extgpmf} is no more directly related to the effective potential which requires that all CG variables are constrained. The GPMF equals the effective potential only in the case where $n=\ncg$.  In view of the physical interpretation of the GPMF along with \tcref{Eq:gpmf} we argue that the GPMF with $n<\ncg$ is the $n$-body approximation to the full GPMF and therefore to the full effective potential.
Being the effective potential $\Veff$, and similarly the PMF, generally a function of all CG variables, is usually compute it in an approximate form \citep{mones2016}, with the most common approach being to assume a multi-body expansion. In practical coarse-graining applications this typically reduces to writing the effective potential as a sum of pairwise and three-body potentials.

One way to achieve this is by employing the quantity $W^{(2)}$, that leads to an extension of the Theorem of Reversible Work \citep{Chandler1987}. Following \cref{eq:extgpmf}, define the 2-body distribution function
\begin{align}\label{eq:reddistr}
	\rho^{(2)}(\bfR_1,\bfR_2) = \ncg(\ncg - 1) \frac{\int \de \bfR_3 \cdots \de \bfR_{\ncg}\int{\de \Aphase{\bfr}  e^{-\beta U(\bfr)}  \prod_{I=1}^{\ncg}\dirac{\Psi_I - \bfR_{I}} }}{\int \de \bfR_1 \cdots \de \bfR_{\ncg}\int{\de \Aphase{\bfr}  e^{-\beta U(\bfr)}  \prod_{I=1}^{\ncg}\dirac{\Psi_I - \bfR_{I}} }}
\end{align}
which represents a joint distribution probability for finding a bead in the position $\bfR_1$ and any other bead at position $\bfR_2$.
From \cref{eq:reddistr}, the Radial Distribution Function (RDF) for a homogeneous and isotropic coarse-grained system can be defined as in \citep{Chandler1987}:
\begin{align}
  g(|\bfR|) = \frac{\rho^{(2)}(0,\bfR)}{\rho^2},\qquad\text{where}\qquad
  \rho := \rho^{(1)}(\bfR_1) = \frac{\ncg}{V},
\end{align}
where $V$ is the volume of the box containing the beads, so in other words, $\rho^{(1)}$ is the
density of beads.

If we sum the forces on the atoms in the bead $1$ averaged over all configurations of the beads $\{3,\ldots,\ncg\}$ keeping fixed the positions of the beads $1$ and $2$ we obtain:
{\small
\begin{align*}
  &\sum_{k=1}^{s_1}\at{{-\frac{1}{s_1c_{1_k}}\pard{W^{(2)}}{\bfr_{1_k}}}}{\bfR_1,\bfR_{2}}
  = - \at{{{\pard{W^{(2)}}{\bfR_1}}}}{\bfR_1,\bfR_2} \\
  &\qquad= \frac{ \int  \de \bfR_3 \cdots \de \bfR_{\ncg}\int{\de \Aphase{} \sum_{k=1}^{s_k}\cip{-\frac{1}{s_1c_{1_k}}\pard{U(\bfr)}{\bfr_k} } e^{-\beta U(\bfr)}  \prod_{I=1}^{\ncg}\dirac{\Psi_I - \bfR_{I}} } }{ \int  \de \bfR_3 \cdots \de \bfR_{\ncg}\int{\de \Aphase{}  e^{-\beta U(\bfr)}  \prod_{I=1}^{\ncg}\dirac{\Psi_I - \bfR_{I}} } } \\
	&\qquad = -\beta^{-1} \frac{\int  \de \bfR_3 \cdots \de \bfR_{\ncg}\int{\de \Aphase{}  e^{-\beta U(\bfr)}  \prod_{I=2}^{\ncg}\dirac{\Psi_I - \bfR_{I}} \sum_{k=1}^{s_1}\cip{\frac{1}{s_1c_{1_k}}\pard{\Psi_1}{\bfr_k}} \pard{}{\Psi_1}\dirac{\Psi_1 - \bfR_{1}} }}{\int  \de \bfR_3 \cdots \de \bfR_{\ncg}\int{\de \Aphase{}  e^{-\beta U(\bfr)}  \prod_{I=1}^{\ncg}\dirac{\Psi_I - \bfR_{I}} } } \\
	&\qquad = \beta^{-1} \pard{}{\bfR_1} \ln \int  \de \bfR_3 \cdots \de \bfR_{\ncg}\int{\de \Aphase{}  e^{-\beta U(\bfr)}  \prod_{I=1}^{\ncg}\dirac{\Psi_I - \bfR_{I}} }  \\
	&\qquad = \beta^{-1} \pard{}{\bfR_1} \ln \cip{\frac{\ncg(\ncg-1)\int  \de \bfR_3 \cdots \de \bfR_{\ncg}\int{\de \Aphase{}  e^{-\beta U(\bfr)}  \prod_{I=1}^{\ncg}\dirac{\Psi_I - \bfR_{I}} }}{\int \de \bfR_1 \cdots \de \bfR_{\ncg}\int{\de \Aphase{}  e^{-\beta U(\bfr)}  \prod_{I=1}^{\ncg}\dirac{\Psi_I - \bfR_{I}} }}} 
	 = \beta^{-1} \pard{}{\bfR_1} \ln g(|\bfR|) 
\end{align*}
}
where we used the properties of the Dirac delta and the fact that $\sum_{i=1}^{s_I}\frac{1}{s_
Ic_{I_k}}\pard{\Psi_I}{\bfr_i}=1$ (which follows from \cref{Eq:constr1}). 

The expression obtained above shows interesting connection with some CG techniques.
The Iterative Boltzmann Inversion (IBI) algorithm \tcitep{Reith2003} is based on the Henderson theorem \tcitep{Henderson1974} that relates the radial distribution function with the pair potential that reproduces this RDF.
The IBI algorithm calculates coarse-grained potential by refining a potential function obtained as an inversion of the expression of the potential of mean force for two beads. The iterations usually starts with the following initial guess: 
\begin{equation}\label{Eq:IBIiter}
	u_0 = -\frac{1}{\beta}\ln{g} \,\,\, \mbox{and}\,\,\, u_{k+1} = u_k + \frac{1}{\beta}\ln\cip{\frac{g_k}{g}}
\end{equation}
where $g_k$ is the radial distribution function obtained at the $k$-th iteration and $u_0$ is the potential at the $0$-th iteration, which in our language represents the two-body approximation of the effective potential $\Veff$. It was shown \tcitep{Hanke2017}  in the context of the fixed point iteration theory, that iteration in \tcref{Eq:IBIiter} is well defined. \tCitet{Hanke2017}  proved that  $\tnorm{u_{k+1} - u}_{{\mathcal{V}_u}} \leq C \tnorm{u_{k} - u}_{{\mathcal{V}_u}}$ if $u_k$ is sufficiently close to $u$, where $C > 0$, $u$ is the true pair potential and the norm $\tnorm{\cdot}_{{\mathcal{V}_u}}$ is defined over the Banach space ${\mathcal{V}_u}$ of the perturbations of the potential $u$. 
The initial iteration of the IBI technique therefore uses the quantity $W^{(2)}(\bfr)$ as first guess. As we have previously identified, this quantity is an approximation of the effective potential. It can be shown that this potential of mean force is a Lennard-Jones type potential and for this reason the first iteration in IBI technique is well defined.
We conclude this section by considering another method used to derive effective CG interaction, the Relative Entropy (RE) framework introduced by \tCitet{ScottShell2008}. This is based on using the Kullback-Leibler divergence \tcitep{Kullback1951} to define a distance metric to estimate how ``close'' two probability distributions are. Given a parameter-dependent CG configurational distribution, the RE method defines the effective CG interactions by minimising the distance, i.e. the entropy, between the CG distribution and the real FG distribution.
Relative entropy has the intuitive property that it is zero when the two distributions represent the same one. The CG interaction that minimises the relative entropy is exactly the effective potential $\Veff$ \tcitep{Chaimovic2011}, which arises naturally from the MZ projection techniques, so that the RE framework is another way to obtain the $\Veff$ potential. It was shown \tcitep{Chaimovic2011} that the IBI and Force Matching methods (which we briefly discussed in \tcref{Sec:EffInt}) are equivalent within the Relative Entropy framework. As we have argued, this result is expected, since the main objective of all these methods is to find a suitable approximation to the effective potential.
We want to highlight here that, in the systematic bottom-up CG with MZ, the best CG approximation is given by the effective potential $\Veff$ and a reliable coarse-grained model crucially requires a good approximation of this quantity, together with the additional fluctuating and dissipative/memory terms. 
The importance of $\Veff$ lies in the fact that it formally represents the exact projected dynamics derived from \textit{first principles} without any calibration. As we have seen above, several methods have been proposed to compute alternative forms of the effective potentials, mostly relying on \emph{a-posteriori} efficient calibration against reference FG simulations. Therefore, all the other approximations (e.g., neglecting memory and fluctuating terms, sampling errors, time-stepping errors, etc.) are implicitly affecting this procedure, and the resulting CG models are guaranteed  to recover only the  quantities and scenarios used for calibration. For other observables, instead, a quantitative estimation of the errors is not generally available. This is only partially overcome by the MZ approach which however relies on computing additional terms.

\subsection{Hybrid systems and the virtual sites approximation}

\noindent To simplify the implementation of atom-bead interactions, we now review a method recently used by \citet{DiPasquale2012a}. The expression in
\cref{Eq:ForceAt} represents the force acting on each atom in the hybrid system due to all the other atoms that are being coarse-grained. In hybrid systems, it may be that atoms which are not coarse-grained into beads can nevertheless be assigned a \textit{virtual bead}.

Supposing this procedure is carried out, then in accordance with the definitions given in \cref{Sec:Def}, we introduce the total number of the virtual sites in the system $\nvs$, which is the total number of CG and virtual beads, and in analogy with the procedure laid out in \cref{Sec:Def}, set
\begin{align}
	S_\mu^{VS}=\{I \in \{1,\ldots,\ncg\}\,\; \mbox{such that}\,\;I\,\;\mbox{is and atom and}\,\,I\,\,\mbox{is in the}\,\; & \mu\mbox{th virtual site} \}, \nonumber \\ 
	&\,\mbox{and} \,\, \mu \in \{1,\ldots,\nvs\}.
\end{align}
The virtual site notation uses Greek letters to highlight the fact that virtual site index is not the same as the CG index. We assume that all real and virtual beads are disjoint, and the number of elements in each set of specific atoms $S_\mu^{VS}$ will be indicated as $\sigma_\mu=\#S_\mu^{VS}$.

Defining virtual sites allows us to obtain a simpler form for the mixed interaction between atoms and beads.
Summing all forces acting on all atoms belonging to a virtual bead \cref{Eq:ForceAt}, divided by $c_{\mu_k}\sigma_\mu$, we obtain:
\begin{align}\label{Eq:VS}
	\sum_{i=1}^{\sigma_\mu}{-\pard{\Veff}{\bfR_{\mu_i}}\frac{1}{c_{\mu_i}\sigma_\mu}} & =
	 {\aver{\sum_{i=1}^{\sigma_{K}}-\pard{ \pot^{AB}}{\bfr_{\mu_i}} \frac{1}{c_{\mu_i}\sigma_\mu}}_{\Bphase{}} } 
\end{align}
where in the above equation we used the linearity of the projection $\pproj$.
Comparing with \cref{Eq:VeffB}, the forces arising from other beads acts on the group of atoms as if they were a proper bead. A virtual site represents the expected location of the atom inside a virtual site had they been coarse-grained. In such a system we will write the interaction between beads and virtual sites as:
\begin{equation}
		-\pard{V^{\mbox{eff}}}{\bfR^{VS}_{\mu}} = \aver{\sum_{i=1}^{\sigma_{\mu}}{\cip{-\pard{\pot^{AB}} {\bfr_{\mu_i}}\frac{1}{c_{\mu_i}\sigma_\mu}}} }_{\Bphase{}} = \aver{\bfF^{VS}_{\mu}}_{\Bphase{}}
\end{equation}
In the above equation we wrote a way to average the force on a group of atoms in the same way used for beads. The only difference with \cref{Eq:VeffB} is the presence of the contribution of the potential term involving the interactions among beads and atoms $\pot^{AB}$ only.  
The importance of this identification lies in the fact that, if the virtual beads are identical to
a subset of the true beads, there is no need to perform additional calculation.
Once the interaction between beads is obtained, we need a way to map the forces between beads back to atoms forming virtual beads. In this case we can use again the mapping operator for the virtual site  $\Psi^{VS}_{\mu}(\bfr)$ obtaining for the $k$th atom inside the virtual bead
\begin{align}\label{Eq:VSNB}
	-\pard{\Veffs{}}{\bfR_{\mu_k}} = -\pard{\Veffs{}}{\bfR^{VS}_{\mu}}\pard{\bfR^{VS}_{\mu}}{\bfR_{\mu_k}} = -\pard{\Veffs{}}{\bfR^{VS}_{\mu}}\pard{\Psi^{VS}_{\mu}(\bfr)}{\bfr_{\mu_k}} = c_{\mu_k}\aver{\bfF^{VS}_{\mu}}_{\Bphase{}}.
\end{align}
The force acting on the $k$th atom belonging to the $\mu$th virtual site is the force acting on the virtual site weighted by the coefficient $c_{\mu_k}$ which depends on the mapping operator.
The result shown in \cref{Eq:VSNB} was used in \citep{DiPasquale2012a} without clear justification other that the \emph{a posteriori} correctness of final results. Here, it is shown how this assumption can be justified within the MZ framework.
The virtual sites approach avoids the computation of an extra interaction between beads and atoms
by using the interaction between beads already computed, thus reducing the computational overhead.

\subsection{Markovian approximation of the Memory Kernel}\label{Sec:MarkovApprox}

\noindent A good description of the memory term represents probably the biggest challenge in the derivation of the correct dynamic for a coarse-grained model. The problems faced in order to give a proper description of this term are twofold: the calculation of the friction matrix, which in its formal derivation depends from all the fast degrees of freedom coarse-grained away (i.e. the orthogonal dynamics, see \tcref{Eq:fricmatrix} and \tcref{Sec:OrthD}) and the time integral of which the friction matrix is one of the arguments, that needs to be evaluated at each time step (see second term on the RHS \tcref{eq:mz}). Different ways to consider the friction matrix were reported \tcitep{Chorin2002,Kauzlaric2011} as well as to include memory effects in CG simulations, \tcitep{Li2015,Lei2016,Yoshimoto2017,Jung2017}.

One way to reduce the complexity of the equations of motion given in \cref{Eq:DevB} is to approximate the memory kernel by assuming the system is `memoryless', i.e. the fluctuating forces at a given time have no correlation with those at previous times. This assumption is typically valid if the timescale over which the CG variables change is much longer than the timescale over which the fluctuating forces vary \citep{Green1954}. If the characteristic timescale for the evolution of the CG variables is $\tau_R$, and the timescale over which the components of the friction matrix decay is $\tau_D$, then if $\tau_R\gg \tau_D$, then $\bffricmat(t)$ can be considered as `memoryless'.
This is the Markovian approximation, which assumes that
\begin{align}\label{Eq:MarFric1}
	\bffricmat(\Bphase{},t) \approx \tilde{\bffricmat}(\Bphase{})\delta(t).
\end{align}
If $\bffricmat(\Bphase{},t)$ exhibits exponential decay in $t$ at a sufficiently large rate uniformly on the phase space, i.e. $\bffricmat$ takes the form
\begin{equation*}
  \bffricmat(\Bphase{},t) = \lambda\exp{-\lambda t}f(\Bphase{},t)
\end{equation*}
where $f$ is bounded in time,
then the integrals involving the memory kernel may be approximated using Laplace's method as
\begin{multline*}
  \int_0^t{\bffricmat(\Bphase{}(t-s),s)
  \pard{}{\Bphase{}}{\entropy}(\Bphase{}(t-s))}\,\de s
  + \beta^{-1}\int_{0}^{t}{\pard{}{\Bphase{}}\bffricmat\cip{\Bphase{}(t-s),s}}\,\de s\\
   =\bffricmat(\Bphase{}(t),0)
   \pard{}{\Bphase{}}\entropy(\Bphase{}(t))
   + \frac{1}{\beta}{\pard{}{\Bphase{}}\bffricmat\cip{\Bphase{}(t),0}}+ O(\lambda^{-1}).
\end{multline*}
In the above approximation, $\lambda$ should be viewed as a measure of the timescale separation,
i.e. $\lambda=\tau_R/\tau_D\gg 1$.
In this case, $\tilde{\bffricmat}$ is the time independent friction matrix defined as:
\begin{equation}\label{Eq:MarDeltaFric}
  \tilde{\bffricmat}(\Bphase{}) := \bffricmat(\Bphase{},0)
  = \beta\,\aver{\fprojbf_{\Bphase{}}(0,\cdot)\otimes
    \fprojbf_{\Bphase{}}(0,\cdot)}_{\Bphase{}}.
\end{equation}
We note that
\begin{equation}\label{Eq:convFlucF}
  \aver{\fprojbf_I(0,\cdot)\fprojbf_J(0,\cdot)}_{\Bphase{}} = \aver{\Lio_\qproj\bfP_I\otimes \Lio_\qproj\bfP_J}_{\Bphase{}}
  = \aver{\bigg(\frac{\partial \Veff}{\partial \bfR_I}-\sum_{i\in S_I}\frac{\partial U}{\partial \bfr_i}\bigg)\bigg(\frac{\partial \Veff}{\partial \bfR_J}-\sum_{j\in S_J}\frac{\partial U}{\partial \bfr_j}\bigg)}_{\Bphase{}},
\end{equation}
which is a function of $\Bphase{\bfR}$ alone, and therefore derivatives of $\tilde{\fricmat}$
with respect to $\bfP_J$ vanish in \cref{eq:MarkovApprox}. It is also clear that the approximate
friction matrix is symmetric, i.e. $\tilde\fricmat_{IJ} = \tilde\fricmat_{JI}$.

The validity of the Markovian approximation has been repeatedly questioned in literature, in particular for system where the timescale of the sound propagation is comparable with the timescale of the evolution of the CG variables \citep{Cubero2005,Cubero2005a,Cubero2008}. In \citep{Hijon2006,Hijon2008} it was argued that the Markovian approximation may fail to be valid for a chain of oscillators interacting with nonlinear potential, such as a chain of Lennard-Jones particles, but in more complicated systems, such as those treated in more realistic MD simulations, it may be considered
valid.

Assuming the approximation \cref{Eq:MarFric1} is valid, the equations of motion for the atomistic/CG system (\cref{Eq:DevB}) become:
\begin{align}\label{eq:MarkovApprox}
  \pard{\bfP_{K}}{t}
  &= -\omega_{\NA{}K}\pard{\pot^{AA}}{\bfr_{K}} -\pard{\Veff}{\bfR_{K}} - \sum_{J=1}^{\ncg}{{\tilde{\fricmat}_{IJ}}\frac{\bfP_{J}}{M_{J}} }   + \beta^{-1} \sum_{J=1}^{\ncg}  {{ \pard{}{\bfR_{J}}\tilde{\fricmat}_{KJ}(s)}}     + \fprojbf_{K}(t) 
\end{align}

We will make use of the last equation in the following section when we will apply the theory to a simple system.

If, instead of $\tilde{\fricmat}_{IJ}$, we further average over the positions $\Bphase{\bfR}$ of CG particles, $\aver{\cdots}$, we could simplify \cref{eq:MarkovApprox} again, by noticing that $\aver{\tilde{\fricmat}_{KJ}}$ does not depend on
$\bfR$ anymore:
\begin{align}\label{Eq:Langevin}
	\pard{\bfP_{K}}{t} &= -\omega_{\NA{}K}\pard{\pot^{AA}}{\bfr_{K}} -\pard{\Veff}{\bfR_{K}} - \sum_{J=1}^{\ncg}{{\aver{\tilde{\fricmat}_{KJ}}}\frac{\bfP_{J}}{M_{J}} } + \fprojbf_{K}(t).
\end{align}
In this case, we have obtained the usual Langevin dynamics for the momentum, where a friction
constant which is uniform in particle positions is usually assumed.

We want to stress here that we use the Markovian approximation only as a means by which to simplify the calculation; more refined methods to calculate the friction matrix, and we will consider them and consider memory effect exist, and will be considered in future work.

\section{Numerical experiments}\label{Sec:Res}
\noindent Having derived the general structure of equations of motion for a generic CG system, we now numerically investigate the effects of some of the choices and assumptions made when a FG system is coarse-grained, along with their conceptual and practical implications. With this aim in mind, our choice is to consider a simple one-dimensional system with only one type of bead: by studying such a simple system, we hope to better understand which of our assumptions are valid, and which might fail in a more complex system.
Moreover, we focus on the differences between a fully coarse-grained system and its fine-grained version, i.e. we do not consider a hybrid system. Although similar systems have been studied before \citep{Hijon2006}, our aim is to carefully consider and test the errors made in the various approximations required and their parametric dependence in order to better understand their validity.

The particular system we consider is a periodic chain of $\nfg$ atoms of two different species. The two species are assumed to have different masses and stiffnesses, in a repeating pattern: such a system might be viewed as a toy model for a binary alloy. The system is coarse-grained by combining single repeating units into beads. Following the choices made in \cref{sec:analysis}, the CG variables are taken to be the centres of mass of each bead and the corresponding momenta. The number of beads, $\ncg$, will therefore be equal to the number of times the pattern is repeated in the chain. An open-source parallel code, written in Julia~0.6.2, has been developed to solve the full, constrained and coarse-grained dynamics in the model described and, more importantly, to compute all of the components needed for the CG system.
The implementation we have developed allows for a relatively straightforward recalibration of the test system via Julia `types' which are initialised by the user to fix all parameters of the simulation, including the repeating pattern of the beads, the number of beads, and the mass and stiffnesses of the interatomic potentials. Dynamics are implemented using symplectic or pseudo-symplectic schemes to maximise accuracy at low computational cost, and the sampling algorithms employed take advantage of parallelisation to maximise the efficiency of the code.

\begin{figure}[hb]
    \centering
    \includegraphics[width=0.75\textwidth]{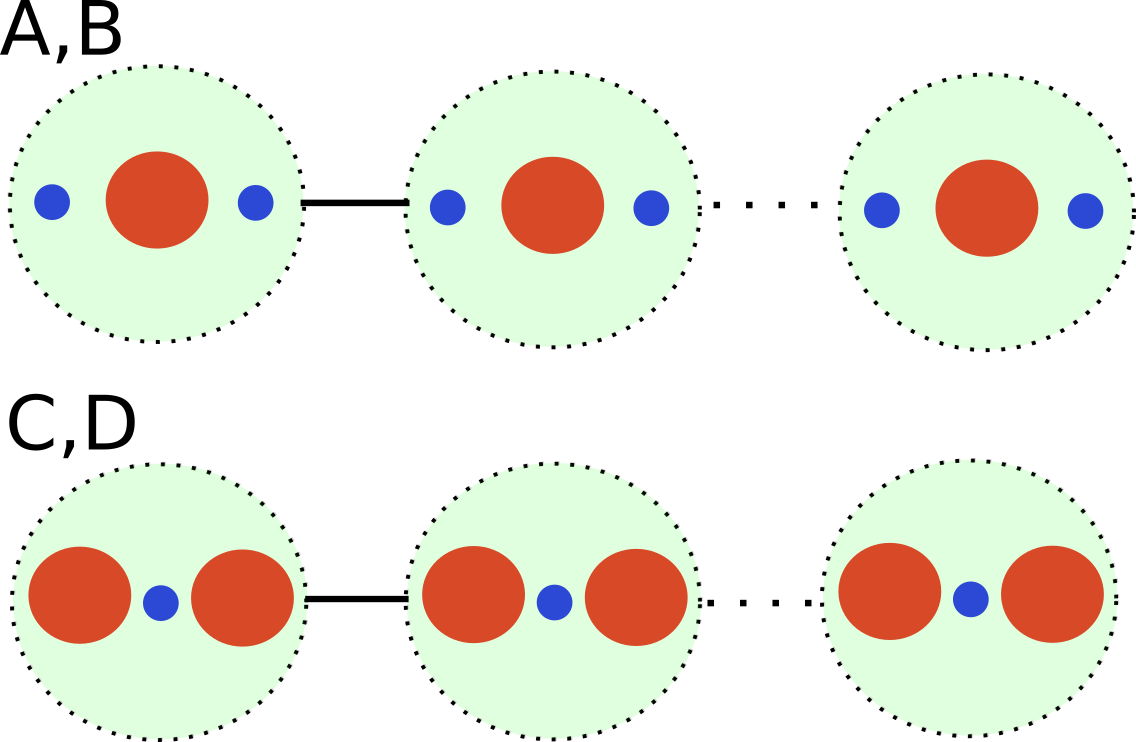}
    \caption{Sketch of the one-dimensional chain and the two chosen configurations. The two species are indicated as red and blue particles. Small blue spheres represent atoms of mass $m$ and big, red spheres represent atoms of mass $M$. The beads are indicated as a dotted line surrounding three atoms.}
    \label{Fig:toyp}
\end{figure}

The particular test cases under study here involve atoms with two different masses, $m=1$ and $M=10$, and the repeating pattern has been chosen to consist of three atoms in two different configurations. We distinguish four cases, denoted as $A$, $B$, $C$ and $D$. In the first two cases, each bead contains two atoms of mass $M$ and a single atom of mass $m$. In the second two cases, each bead is composed of two atoms of mass $m$ and a single bead of mass $M$. The one-dimensional chain in all the configurations considered, along with their coarse-grained version is sketched in \cref{Fig:toyp}. The difference between systems $A$ and $B$, $C$ and $D$ is given by the different interactions among the atoms, as will be explained in detail when we will introduce the specific form of the inter-atomic potential.
Throughout, the number of beads will generally be set to $\ncg=10$; clearly, by fixing the number of beads we fix the number of atoms also, as $\nfg=3\ncg=30$.

The interaction potential between atoms is chosen to be a simple 12-6 Lennard-Jones potential:
\begin{equation}
	\pot_{i,i+1}(r) = 4\epsilon_{i,i+1} \cip{\cip{\frac{\sigma_{i,i+1}}{r}}^{12} - 2\cip{\frac{\sigma_{i,i+1}}{r}}^6}.
\end{equation} 
We choose $\sigma_{i,i+1}=1$ so that the minimum of the potential is attained at $r^*=1$, and the well-depth $\epsilon_{i,i+1}$ will be varied to analyse the effect of the stiffness of the interactions. We choose three values for this parameter $\{0.1,\,1,\,100\}$; the potentials $U_{i,i+1}(r)$ describing the mixed interaction among two different atoms are fixed by choosing the parameters $\epsilon_{i,i+1}$ which obey the Lorentz-Berthelot mixing rule. The sequence of well-depth values for all the system is then given by 
\begin{equation}
  \epsilon_{i,i+1} = \begin{cases}
    \sqrt{\epsilon^a\epsilon^b} & i =1,2,4,5,\dots,\nfg-2,\nfg-1,\\
    \epsilon^a\;\;(\mbox{or}\,\,\epsilon^b) & i=3,6,\dots,\nfg.
  \end{cases}
\end{equation}
where the superscripts $a,\,b$ indicate two different values for $\epsilon$ taken from the set $\{0.1,\,1,\,100\}$.
These choices are paired with the different mass patterns to give 4 distinct systems. The pairs $A$ and $C$, and $B$ and $D$ have the same `internal' interactions (i.e. interactions among atoms belonging to the same bead) and same inter-bead interactions (i.e. interactions among atoms in different beads): a summary of the four cases is reported in \cref{Tab:systems}. In this way, we seek to isolate the effects of the different beads and mapping on the properties investigated.
\begin{table}[h]
	\begin{center}
		\begin{tabular}{ c|c |c |c | c}
	              & \multicolumn{3}{|c|}{Interactions $(\epsilon_{i,i+1})$} & Bead structure  \\
	                                \hline
                   &  $m-M$ & $M-M$ & $m-m$ &\\
                  \hline
    		$A$ & $10$ &  $100$ & $1$  & $m$-$M$-$m$\\ 
   			$B$ & $1$  &  $.1$  & $10$ & $m$-$M$-$m$  \\
  			$C$ & $10$ &  $1$   & $100$ & $M$-$m$-$M$ \\
			$D$ & $1$ &   $10$ & $.1$  & $M$-$m$-$M$
		\end{tabular}	
	\caption{Summary of parameters in the 4 configurations considered.}
	\label{Tab:systems}
	\end{center}
\end{table}

Letting $\bfp_i$ and $\bfr_i$ respectively be the momentum and position of the $i^{\text{th}}$ atom, the Hamiltonian for system $X=A,B,C,D$ reads
\begin{equation}
  \Ham^X(\bfr,\bfp) = \frac12\bfp^T(M^X)^{-1}\bfp +\sum_{i=1}^{\nfg-1}\pot^X_{i,i+1}\cip{\bfr_{i+1}-\bfr_i}+\pot^X_{\nfg,1}(\bfr_1-\bfr_\nfg+r^*\nfg),
  \label{eq:1DHamiltonian}
\end{equation}
where the mass matrix $M^X$ and pair potentials $U^X_{i,i+1}$ have been chosen with the relevant
parameters.
The final term reflects the coupling of the right-hand end of the chain to the left-hand end, where we
have chosen a total chain length of $r^*\nfg=30$; other fixed volumes could be selected by making other choices.

As mentioned above, in all cases, the dynamics were implemented using symplectic schemes, and the
time step chosen was $\Delta t = 10^{-3}$; we note that $\Delta t\leq \min\sqrt{M/U_{i,i+1}''(1)}\approx 5.89\times 10^{-3}$, the shortest timescale in the system. 
Below, we present the results of simulating different dynamics: for convenience, we term these dynamics:
\begin{itemize}
	\item[(FGD)] The \emph{Fine-Grained Dynamics} when solving \cref{eq:hamiltonsyst} with the Hamiltonian defined in \eqref{eq:1DHamiltonian};
	\item[(OD)] the \emph{Orthogonal Dynamics} when solving \cref{Eq:orth};
	\item[(DCGD)] the \emph{Deterministic Coarse-Grained Dynamics} when solving \cref{eq:MarkovApprox} neglecting the dissipative and fluctuating terms; and
	\item[(MMZD)] the \emph{Markovian Mori-Zwanzig Dynamics} when solving the full \cref{eq:MarkovApprox}, with the memory and fluctuating terms replaced by a  position-dependent\footnote{(i.e., fluctuation and dissipation terms that depends on the inter-particle distance)} Langevin dynamics.
\end{itemize}
To numerically integrate (MMZD), we have used a version of the BAOAB integrator \citep{Leimkuhler2013,Leimkuhler2016} adapted to the case of non-constant diffusion.


\subsection{Sampling algorithm}

\noindent In order to compare the predictions of an approximation of the MZ projection to the true dynamics,
we must calculate an approximation of the effective potential, given in \cref{Eq:EntrInt},
 and the friction matrix, as given in \cref{Eq:fricmatrix}. Here we report the algorithm we used to generate the results that will be used for the discussion. The basic idea is to sample using the OD.
 
When the mapping operator onto CG variables is chosen to be a canonical transformation, the OD has the property that it remains Hamiltonian, and therefore preserves energy, sampling an appropriate marginal of the NVE ensemble. Moreover, in the case of a \emph{linear} canonical transformation, the OD takes the explicit form given in \cref{Eq:orth}; these equations may be integrated using a leapfrog scheme, just as for the FGD.

We first fix an initial condition for the simulation, along with the corresponding value for the Hamiltonian $\Ham(\bfp,\bfr)=E$. A trajectory of the FGD is then computed using a leapfrog scheme, and samples of fine-grained positions are recorded. For each sample, CG positions and momenta are extracted, and a longer trajectory of the OD is then run with the sample positions and momenta as initial condition. Along trajectories of the OD, time averages of the first and second moments of the effective forces between beads are stored.
The first quantity gives the value of the force acting between adjacent beads given the positions of the centre of mass, while the latter is used to compute the friction matrix.
Here, we implicitly assume that \eqref{Eq:orth} generates an ergodic dynamics on the relevant marginal of the NVE ensemble; practically speaking, this assumption cannot be verified, but is commonly assumed in MD simulations.

Since the OD sampling step can be run concurrently with no communication between processes, it is ideal for parallelisation: each time average provides a value for the effective force between beads given a distance between centres of mass of adjacent beads. A pseudo-code script for the sampling algorithm used is reported in the SM.

The method we used to calculate the effective potential is very similar to the force-matching method \citep{Izvekov2004}. However, as mentioned above, here we \emph{directly} derive the effective potential from the sampling performed during OD simulations, without prescribing a functional form for the CG potential, and therefore have no need of minimisation techniques.
Using the constrained dynamics \tcref{eq:orthForces}, we sample coefficients of the macroscopic dynamics, using the following algorithm:
\begin{algorithm} [H]
\begin{algorithmic}[1]
  \Procedure{Constrained Sample}{Configuration of atomistic system, target$_1$, target$_2$}
  \State Generate initial condition for fine-grained system
   \While{$\#$ of samples $<$ target$_1$}
   \State Run fine-grained dynamics to generate initial condition for constrained dynamics
   \While{length of constrained dynamics trajectory $<$ target$_2$}
   \State Compute timestep using integrator for constrained dynamics
   \State Update statistical estimators
   \EndWhile
   \EndWhile
\EndProcedure
\end{algorithmic}
\end{algorithm}

\subsection{Data-driven approximation of the effective potential and memory kernel}\label{Sec:ApproxVeff}
 
\noindent Additional assumptions must be made in order to enable practical calculation; we highlight each such assumption in order to understand whether we may consistently extrapolate from the conclusions we draw for this simple system to a more realistic situation.

Firstly, we make an \textit{equivalence assumption} for the interaction between beads, i.e., using the symmetry of the system, we sample interactions between similar beads concurrently. In our case, this assumption is valid as every bead is identical and, since they lie in a periodic arrangement, are equivalent by translation. We can therefore calculate the interaction between any combination of beads in the same arrangement relative to one another and assume that the interaction computed is valid for any equivalent pair in the system. 

Secondly, we make \textit{parametric assumptions}, i.e. we choose the form of the coarse-grained variables, which leads to a choice of parametrisation of the effective potential. The choice of the variables is often intuitive and completely natural (inter-particle distances, angles, etc.), but currently there exist no systematic way (apart from purely data-driven approaches) to find an `optimal' representation of the potential.

In many approaches, a third \textit{functional assumption} is made, in which the functional form of the potential is chosen, and optimised within some parameter space. In the following test case, we make no functional form assumptions, leaving the functional form free, and we therefore avoid any intrinsic bias in our method.

The last and most significant simplification is a \textit{sparsity assumption}: not only do we make a specific choice of coarse-grained variable, we also assume that the effective potential may be approximated as a sum of $n$-body interactions between beads, where $n$ is of a lower dimension than the dimension of the coarse-grained system. In this test case, we compute only 2-body interactions based on the distance to the `nearest-neighbour' beads only, and use this as an
approximation to the full effective potential, neglecting all other variables.
These assumptions mean that we drastically reduce the dimensionality of the space which we need to sample to compute the effective potential, from the total number of coarse-grained variables to simply one, i.e., the inter-bead distance. This inevitably simplifies the exploration of the entire constrained phase space, as well as the representation and interpolation of the approximate potential in the parameter space, and thus improves the computation time required to adequately sample the mean force. Clearly, it is possible, at the expense of further computational effort and more sophisticated interpolation methods, to improve the effective potential by accounting for multi-body interactions.

It is important to highlight here that, although in most previous studies these approximations were not explicitly identified, they are crucial to make any coarse-graining method applicable. In principle, any of the above assumptions could cause a significant deterioration in the accuracy and effectiveness of a systematic MZ projection approach. For example, in realistic systems, even chemically identical beads do not fully satisfy the equivalence assumption as they occupy different relative positions within a larger molecule. An overly sparse approximation of the effective potential could fail to appropriately penalise certain configurations, resulting in a molecule which is too flexible, or fluctuates too much.
\begin{figure}[t]
    \centering
    \includegraphics[width=0.75\textwidth]{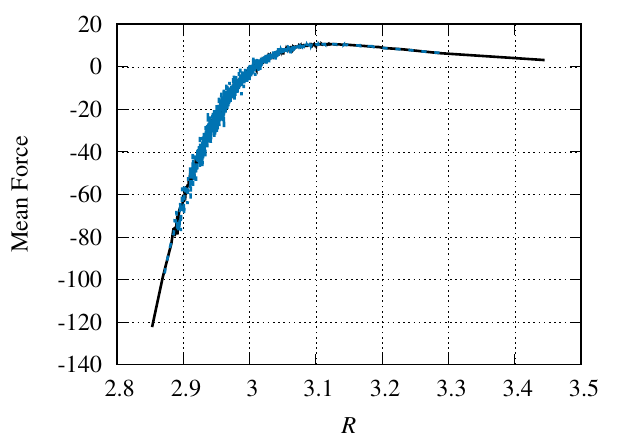}
    \caption{Plot of the mean force for the system $C$. Mean force calculated with all the beads constrained (dotted blue line), and mean force calculated with a single bead constrained while the rest of the system evolve with the full dynamics (solid black line).}
    \label{Fig:FFcompare}
\end{figure}
In spite of the caveats above, we are able to illustrate the validity of both the equivalence and sparsity assumptions in this test case. In \cref{Fig:FFcompare}, the mean force for system $C$ is shown, calculated by constraining a single pair of beads; superimposed are samples of the mean force between two particles calculated by constraining all beads in the system. The pair potential clearly accounts for the majority of the effective force, while significantly reducing sampling error, and in this case therefore appears a reasonable and statistically-stable approximation.

\begin{figure}[t]
    \centering
    \includegraphics[width=0.75\textwidth]{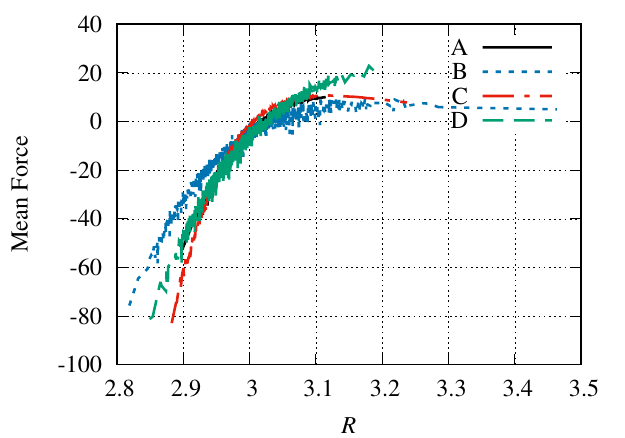}
    \caption{Plot of the mean force for the four systems considered: system $A$ (black solid line), $B$ (dotted blue line),  $C$ (red dashed-dotted line), $D$ (dashed green line). The black line is partially covered by, and mostly overlapping with the blue ones.}
    \label{Fig:MeanF}
\end{figure}

In \cref{Fig:MeanF}, we report the results of the computation of the mean force between beads computed via OD simulations in each of the four test cases $A$-$D$.
The main behaviour of each curve depends from the parameters used in the description of the interactions (see \cref{Tab:systems}), however we can find some expected similarities. In each case, we observe that the equilibrium bond length lies close to $R=3$, which corresponds to both the minimum of the $U_{i,i+1}$ potential for three consecutive atoms, as well as the value expected due to the volume constraint: changing the density of the system would result in different minima for the effective potential.

\begin{figure}[ht]
    \centering
    \includegraphics[width=0.75\textwidth]{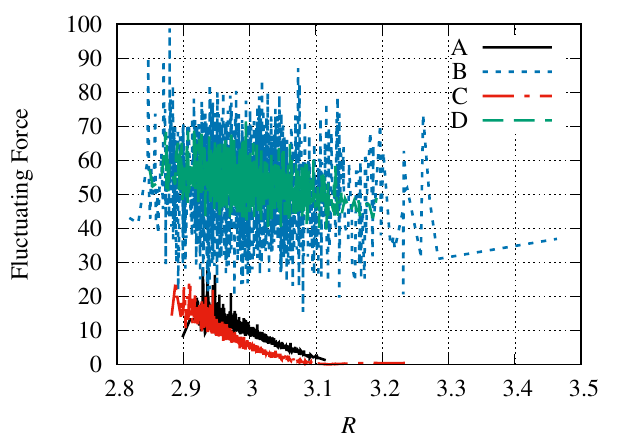}
    \caption{Plot of the fluctuating force for the four systems considered: system $A$ (black solid line), $B$ (dotted blue line),  $C$ (red dashed-dotted line), $D$ (dashed green line).}
    \label{Fig:FlucF}
\end{figure}
While the mean force observed for the four systems are qualitatively similar, the fluctuating part of the force exhibits more significant differences, shown in \cref{Fig:FlucF}. Even in such a simple system, all the four chains $A$-$D$ have a clear dependence of the fluctuations on the inter-bead distance $R$, demonstrating that a simple Langevin dynamics (with white noise fluctuations and dissipation having a magnitude independent of the state of the system) is not an adequate approximation.
The fluctuations for cases $B$ and $D$ are noisier and of a much greater magnitude than that in systems $A$ and $C$, which appears to be a reflection of the increased stiffness of the bonds between beads in the latter cases. This would seem to result in greater fluctuations in the force for similar interatomic displacements, and leads us to hypothesis that the mixing timescale for the OD may be determined by the bead mass divided by the stiffness of inter-bead interactions. It is interesting to notice that, despite these differences, the corresponding effective potentials look instead very similar. This means that, in practical applications, even when the effective potential is easily and stably computed, significant dissipative effects might be present that are not visible in the deterministic effective potential.
Another interpretation of the presence of larger fluctuations in the cases $B$ and $D$ is that we have made a `worse' choice of mapping, and parametrisation of the potential; in view of \cref{eq:mz}, a memory of larger magnitude means the effective potential accounts less of the force felt by the beads.

The integration of the mean force (\cref{Fig:MeanF}) with respect the distance $R$ between the beads leads to the effective potential $\Veff$. 
\begin{figure}[h]
    \centering
    \includegraphics[width=0.75\textwidth]{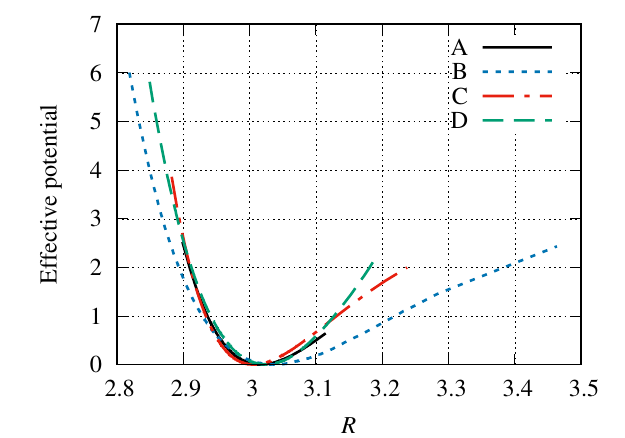}
    \caption{Plot of the effective potential for the four systems considered: system $A$ (black solid line), $B$ (dotted blue line),  $C$ (red dashed-dotted line), $D$ (dashed green line).}
    \label{Fig:Veff}
\end{figure}

In \cref{Fig:Veff}, the effective potential is plotted: we note that oscillations in the mean force are averaged through by the integration. In practice, we use a kriging-type interpolation algorithm \citep{Jones2001} to obtain a smooth approximation of the mean force directly from the values shown in \cref{Fig:MeanF}. This approach allows us to compute forces at arbitrary values of the inter-bead distance, and simultaneously filters the noise resulting from sampling error. Details of this algorithm are discussed further in the SM.

If we assume that the forces on bead $I$ can be decomposed as independent `stress' contributions
$\sigma_{I+1,I}$ and $-\sigma_{I,I-1}$, due to the interaction with bead $I-1$ and bead
$I+1$, then
\begin{align*}
  \fricmat_{II}(\bfR) = \mathbb{E}[\fprojbf_I\otimes\fprojbf_I|\bfR]
  &=\mathbb{E}[(\sigma_{I+1,I}-\sigma_{I,I-1})\otimes(\sigma_{I+1,I}-\sigma_{I,I-1})|\bfR]\\
  &=\mathbb{E}[\sigma_{I+1,I}\otimes\sigma_{I+1,I}|\bfR]+\mathbb{E}[\sigma_{I,I-1}\otimes\sigma_{I,I-1}|\bfR];
\end{align*}
on the latter line we have assumed that the cross terms are negligible, i.e.:
\begin{equation*}
  \mathbb{E}[\sigma_{I+1,I}\otimes\sigma_{I,I-1}|\bfR],\,\mathbb{E}[\sigma_{I,I-1}\otimes\sigma_{I+1,I}|\bfR]\approx 0.
\end{equation*}
The equivalence between the projection and an expected value is discussed in the SM.
This assumption is natural in view of the hypothesis that pair interactions are sufficient to
capture the behaviour of the system, and thus fluctuating forces between adjacent beads are not
correlated. Similarly, we obtain
\begin{align*}
  \fricmat_{I,I+1}(\bfR) = \mathbb{E}[\fprojbf_I\otimes\fprojbf_{I+1}|\bfR]
  &=\mathbb{E}[(\sigma_{I+1,I}-\sigma_{I,I-1})\otimes(\sigma_{I+2,I+1}-\sigma_{I+1,I})|\bfR]\\
  &=-\mathbb{E}[\sigma_{I+1,I}\otimes\sigma_{I+1,I}|\bfR].
\end{align*}

Given the interparticle distances $\bfR_{I+1}-\bfR_I$, we compute a smooth approximation of the
variance of the fluctuating force between beads, i.e.
\begin{equation*}
  \gamma_{I+1,I} := \mathbb{E}[\sigma_{I+1,I}\otimes\sigma_{I+1,I}|\bfR]
\end{equation*}
again using the same kriging-type interpolation algorithm as for the effective potential. We then define the
tridiagonal matrix
\begin{equation}\label{Eq:TridMat}
  \Sigma_{IJ} = \begin{cases}
    -\gamma_{I+1,I} & I=J-1\mod \ncg,\\
    \gamma_{I+1,I}+\gamma_{I,I-1} & I=J,\\
    -\gamma_{I,I-1} & I=J+1\mod \ncg.
  \end{cases}
\end{equation}
The fluctuating forces and memory terms may then be generated together by
computing a realisation of the stochastic dynamics
\begin{equation}
  -\tfrac12\Sigma(\bfR_t) M^{-1}\bfP_t\,\de t+\Sigma(\bfR_t)^{1/2}\,\de\mathbf{W}_t,
\end{equation}
where $\Sigma(\bfR)^{1/2}$ is the positive definite square root of the matrix $\Sigma(\bfR)$, and $M$ is the matrix of bead masses, and $\mathrm{d}\mathbf{W}_t$ denotes an increment of a Brownian motion in $\mathbb{R}^{\ncg}$. In practice, a realisation of this dynamics is computed in a similar way to the BAOAB scheme described in \citep{Leimkuhler2013}.

\subsection{Timescale separation and the validity of the memory approximation}

\noindent Sampling the fluctuating force allows us to calculate the components of the friction matrix by using \cref{Eq:MarDeltaFric} that will be used to simulate the MMZD.
The usual physical justification for \cref{Eq:MarDeltaFric} relies on the assumption that the decay of the time of the correlation of the fluctuating force is much smaller that the relaxation time of the system \citep{Hijon2009}. These two quantities can be estimated by using the fluctuating force auto-correlation function (FFACF) and the velocity auto-correlation functions for the beads (VACF)  calculated via a FGD simulation.

\begin{figure}[ht]
\makebox[\textwidth][c]{
    \includegraphics[width=0.5\textwidth]{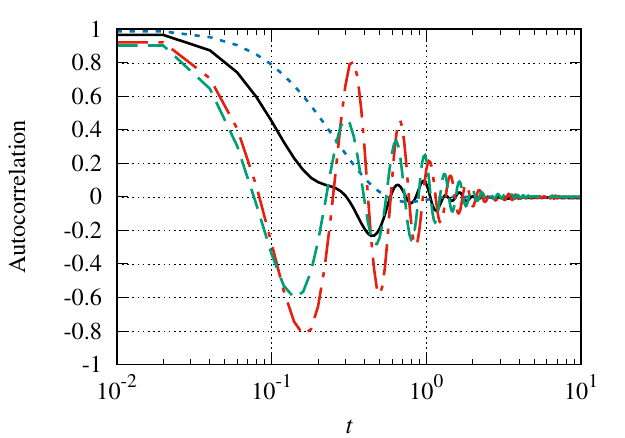}
    \includegraphics[width=0.5\textwidth]{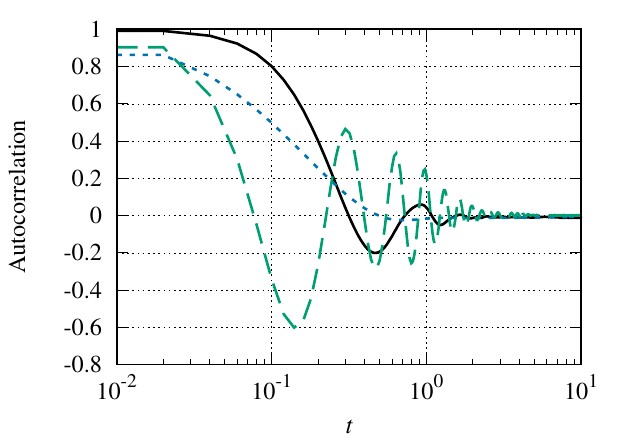}
    }
    \caption{Time autocorrelation function (ACF) of forces and momenta. In the left panel the ACF of the forces in the FGD (black solid line) is compared with that of the forces in the OD (red dot-dashed line), the MMZD (blue dashed line), and the `orthogonal' component of the force in the FGD (green dashed line).
    In the right panel, the ACF of the momenta in the FGD (black solid line) is compared with the ACF of the momenta in the MMZD (blue dashed line), and the ACF of the momenta in the OD (green dashed line). All calculations refer to  the system $C$ (see \cref{Tab:systems}) and are averaged in time and over all the beads.}
    \label{Fig:autocorr}
  \end{figure}

These autocorrelation functions for the model for the system $C$ are plotted in \cref{Fig:autocorr}.

The results are distinct from those shown in \citep{Hijon2006} for a chain of Lennard-Jones particles, as the timescale separation is less distinct. According to the usual rule-of-thumb, which states that the velocity autocorrelation function should decay more slowly than the autocorrelation function of the fluctuating forces, it should follow that there is no clear scale-separation which fully justifies the Markovian approximation. Possible explanations for the lack of scale separation are that the system analysed here is much smaller than system shown in \citep{Hijon2006}, and the bead size chosen here is also much smaller.

  Nevertheless, we note that the autocorrelations of both forces and momenta in the MMZD   (represented by blue dashed curves in \cref{Fig:autocorr}) lack oscillations, and indeed, this   agrees with the significantly weaker oscillations observed in the autocorrelations computed   from the FGD (shown in black). This suggests that while a distinct timescale separation is not
  evident from the autocorrelation function, there may be a `homogenization' effect due to the   rapid oscillation of the fluctuating forces, and hence the Markovian approximation remains an  appropriate first estimation of the memory effect, even in the present case.

\subsection{The effect of the memory approximation on observed system properties}

\noindent After obtaining an approximation of both the effective forces and the fluctuations, we compare the benefits of simulating the MMZD over the DCGD (i.e. dynamics evolving under the effective potential only, ignoring the memory terms), using as a target results obtained from the FGD. The observables chosen for comparison are the distribution of distances between centres of mass of adjacent beads, the momentum distribution of a single bead, and the evolution of the mean-squared displacement between the centres of mass of two adjacent beads over time. The two former observables are linear spatial statistics, and as such, it might be expected that either dynamics would capture them well.
The latter statistic is a dynamical property of the system, and is an average of a nonlinear function, so we would expect to see non-trivial differences between the observables when computed during a simulation of the DCGD and the MMZD.

\begin{figure}[t]
\makebox[\textwidth][c]{
    \includegraphics[width=0.5\textwidth]{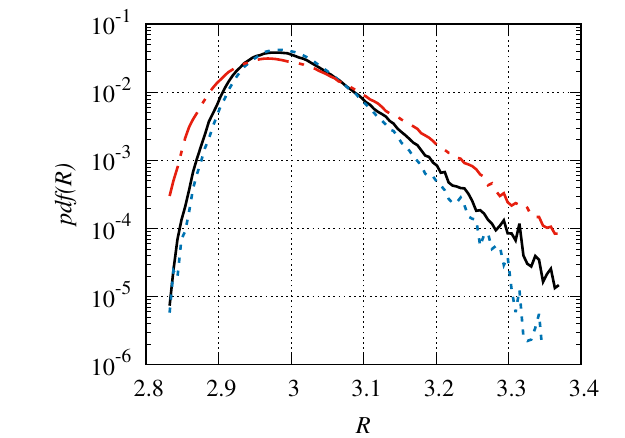}
    \includegraphics[width=0.5\textwidth]{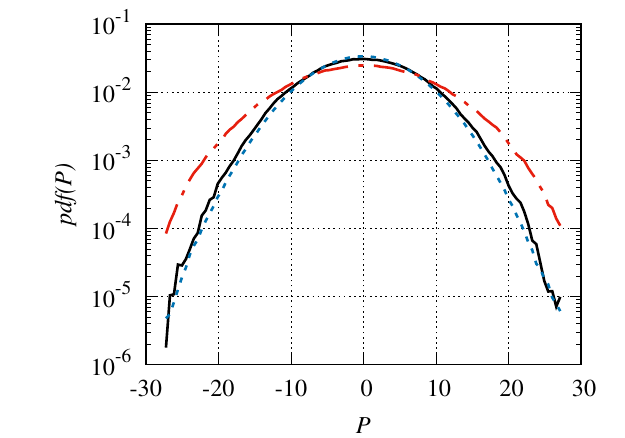}
    }
    \caption{PDFs of inter-bead distances and bead momenta. Comparison between the Fine-Grained dynamics (black solid line), the Deterministic Coarse-Grained dynamics (red dot-dashed line), and the Markov-Mori-Zwanzig dynamics (blue dashed line). All calculations refer to system C (see \cref{Tab:systems}).}
    \label{Fig:compareDistr}
\end{figure}

In \cref{Fig:compareDistr}, we report the distributions of the momenta and positions obtained through simulation of the FGD (shown in black), as well as the DCGD using only the effective potential (shown in red), and the MMZD \cref{Eq:DevB} (shown in blue). The addition of the fluctuating force correction clearly improves the distributions, and indeed captures the tail behaviour of the distributions surprisingly accurately, in view of the relatively crude approximation of the memory term using \cref{Eq:MarDeltaFric}. In all cases, the mean distance between particles obtained is 3.00 to three significant figures. However, it is clear that even in this extremely simple toy system, the addition of the fluctuating terms does indeed improve the accuracy of the approximation in the case of these distributions. %

The most important result shown in \cref{Fig:compareDistr} is the behaviour of the curve for the Markov-Mori-Zwanzig dynamics (dashed blue curve) relative to the curve for the Deterministic Coarse-Grained dynamics (red dot-dashed line). A CG potential obtained from some static distribution of the FG system (e.g. potentials obtained via IBI which are based on radial distribution functions) require an  optimization, usually represented by a series of iterations to recover the correct distribution. In our case, we performed no optimisation of the potential: through sampling, we have simply approximated all of the terms in \cref{eq:MarkovApprox}. In particular, we can see that the main effect of the addition of the fluctuation-dissipation terms is to reduce the width of the distributions: while the average is satisfactorily captured by the DCGD (as shown by the overlapping peaks of the distributions), the tails are not correctly described. DCGD predicts a higher variance of bead velocities, pushing the beads further from equilibrium than they are in the real simulation. The presence of the additional terms provides improved variance without the need for optimisation of the effective potential.  
The only significant difference between the distributions of the positions of the MMZD and the FGD is the right tail of the position distribution; the region around the equilibrium position ($R \in [2.85,3.2]$) is relatively accurate, but away from the equilibrium position the two distributions start to diverge; this difference may be either due to insufficient sampling, or simply because the Markovian approximation is insufficient in these regions: improving the approximation would inevitably improve the description of the system.

\begin{figure}[ht]
\makebox[\textwidth][c]{
    \includegraphics[width=0.5\textwidth]{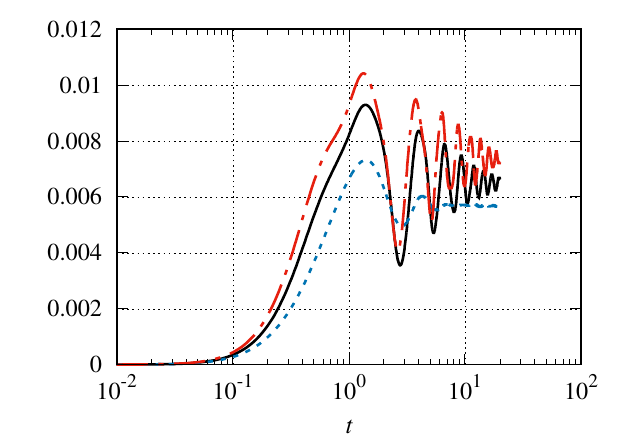}
    }
    \caption{Mean-squared displacement (MSD) of the beads. Comparison between the Fine-Grained dynamics (black solid line), the Deterministic Coarse-Grained dynamics (red dot-dashed line), and the Markov-Mori-Zwanzig dynamics (blue dashed line). All calculations refer to system C (see \cref{Tab:systems}).}
    \label{Fig:compareMSD}
\end{figure}
In \cref{Fig:compareMSD}, we plot the mean-squared displacement (MSD) of the inter-bead distance in time for the MMZD, DCGD and FGD, showing the effect of the different approximations on the dynamics of the beads. As expected, the particles simulated under the DCGD move faster than in the FGD, whereas the opposite is true for the MMZD. In this case we do not observe as good an agreement between the FGD and the MMZD. This is likely to be due to the crudeness of the Markovian approximation, as discussed above. In particular, the results seem to indicate that the value of the friction matrix is too large: each peak of the blue curve lies beneath the black one. The same seems true in view of the fact that the autocorrelation of the momenta appears to decay too rapidly, as seen in the right-hand side of \cref{Fig:autocorr}.

The usual way to deal with the difference in the dynamical quantities between CG and FG simulations (e.g. diffusion coefficient, MSD) is to include into the CG model a way to map the time scales span by the FG system into the CG one \tcitep{Fritz2011}. The time mapping consists in a scaling of the dynamic quantities obtained by CG simulations by a suitable parameter that measure how faster the CG dynamics is with respect the underlying FG system one \tcitep{Harmandaris2009,Fritz2009a,Depa2007}. In the framework of the MZ theory the same results can be achieved by using better defined quantities which can be formally derived from first principles. A more refined description of those is therefore essential for a better description of CG models.

\section{Conclusion}
\label{Sec:Conclusion}

While the development of the Mori-Zwanzig (MZ) theory in the context of non-equilibrium thermodynamics dates back to 1961, only in recent years has it been developed as a theoretical framework in which to systematically derive Coarse-Grained models that might be applicable to Molecular Dynamics. The first purpose of this work has been to give an accessible, rigorous derivation of a class of CG models through the MZ formalism. We recalled the theory, giving a full derivation in the SM, and provided a sketch of the most crucial points in the derivation of the various equations within the MZ framework. 

We then discussed the relationship between the MZ approach and other coarse-graining approaches used in practice, including various methods of computing effective potentials, and the Virtual Sites approximation \citep{DiPasquale2012a}. In the latter case, the MZ formalism provides a justification of the approach, previously used on an \emph{ad-hoc} basis and justified retroactively. More generally, it can provide ways to assess \emph{a-posteriori} the validity
of coarse-graining approaches, and indeed provides a framework for quantifying their approximation errors.

Finally, we numerically studied a toy model, with which we illustrated some features of the MZ derivation. An open source code to efficiently compute the effective potential, friction matrix and fluctuating terms was written in Julia. We consider four systems with different parameter, and in all cases, no significant scale separation between the dynamics of the beads and individual atoms was observed. Nevertheless, a comparison was made between the true and approximate dynamics, and as
expected, systems with `more rigid' beads, having greater internal stiffness, showed much weaker memory effects.
In all cases, we observed that a simple Langevin dynamics (with a friction coefficient independent of position) does not appear to be appropriate for generic CG systems.

Despite the lack of a clear scale separation, the Markovian approximation we used, was able to represent very accurately the equilibrium distribution of positions and momenta, while a deterministic CG fails to be as accurate, although neither approach succeeded in accurately capturing the dynamical observable we chose to test, suggesting that oversimplifying the memory means that fluctuations in the autocorrelation of forces are not accurately reproduced.

The projection operator we considered in this work is applied uniformly in the space of the FG variables to derive the CG model. However, an extension of the projection operator was considered in the development of H-ADDRESS  \tcitep{Potestio2013} where the projector in applied to an open subspace of the FG domain \tcitep{DelleSite2017}. The latter also show how powerful is the MZ theory in the framework of the coarse-grained simulations and the needs to be further and more thoroughly investigated.  

\bibliographystyle{apsrev4-1}
\bibliography{bibliography}

\end{document}